\documentclass[aps,prd,longbibliography,twocolumn,preprintnumbers,nofootinbib,
superscriptaddress,amsmath,floatfix]{revtex4}

\usepackage{amssymb}  
\usepackage{graphicx,subfigure}
\usepackage{exscale}
\usepackage{bbold}
\usepackage{textcomp}
\usepackage{enumerate}
\usepackage [english]{babel}
\usepackage[utf8]{inputenc}
\usepackage [autostyle, english = american]{csquotes}
\usepackage{hyperref}
\MakeOuterQuote{"}

\usepackage{color}

\usepackage[normalem]{ulem}  

\newcommand{\non}{\nonumber\\}

\newcommand{\be}{\begin{equation}}
\newcommand{\ee}{\end{equation}}
\newcommand{\bea}{\begin{eqnarray}}
\newcommand{\eea}{\end{eqnarray}}
\newcommand{\ba}[1]{\begin{array}{#1}}
\newcommand{\ea}{\end{array}}

\newcommand{\MKK}{M_{\mathrm{KK}}}

\begin{document}

\title{Building a realistic neutron star from holography}

\author{Nicolas Kovensky}
\email{nicolas.kovensky@ipht.fr}
\affiliation{Institut de Physique Th\'eorique, Universit\'e Paris Saclay, CEA, CNRS, Orme des Merisiers, 91191 Gif-sur-Yvette CEDEX, France.}

\author{Aaron Poole}
\email{a.d.poole@soton.ac.uk}
\affiliation{Mathematical Sciences and STAG Research Centre, University of Southampton, Southampton SO17 1BJ, United Kingdom}

\author{Andreas Schmitt}
\email{a.schmitt@soton.ac.uk}
\affiliation{Mathematical Sciences and STAG Research Centre, University of Southampton, Southampton SO17 1BJ, United Kingdom}

\date{28 February 2022}

\begin{abstract} 
We employ the recently improved description of dense baryonic matter within the Witten-Sakai-Sugimoto model to construct neutron stars. In contrast to previous holographic approaches, the presence of an isospin asymmetry allows us to implement 
beta equilibrium and electric charge neutrality. As a consequence, we are able to model the crust of the star within the same
formalism and compute the location of the crust-core transition dynamically. 
After showing that a simple pointlike approximation for the baryons fails to satisfy astrophysical constraints, we demonstrate that our improved description 
does account for neutron stars that meet the current experimental constraints for mass, radius, and tidal deformability. However, we also point out tensions in the parameter fit and large-$N_c$ artifacts and discuss how to potentially resolve them in the future. 
\end{abstract}

\maketitle


\section{Introduction}

Dense matter in the interior of a neutron star is notoriously difficult to understand from first principles. The strong coupling nature of the problem -- 
neutron star matter is dense, but not asymptotically dense -- makes the 
AdS/CFT correspondence ("holography") \cite{Maldacena:1997re,Witten:1998zw} a viable theoretical tool. While the holographic principle allows for a rigorous strong-coupling calculation, 
studies are currently and for the foreseeable future constrained to string models whose dual is more or less different from the relevant underlying field theory, Quantum Chromodynamics (QCD). Moreover, at least in the most accessible approximations, the results are strictly valid only in the strong-coupling limit and for a large number of colors $N_c$ of the dual field theory, while $N_c=3$ in QCD.  Nevertheless, for instance in the context of the quark-gluon plasma in heavy-ion collisions \cite{Casalderrey-Solana:2011dxg} or deep inelastic scattering in the so-called small-$x$ regime \cite{Brower:2010wf,Kovensky:2018xxa}, considerable qualitative, and in some cases quantitative, progress has been made with the help of holography, in conjunction with more traditional field-theoretical approaches. 

Recently, holographic models have also increasingly been used to understand the regime of large baryon densities and the properties of compact stars. Several studies use holography for a description of quark matter, and combine it with "ordinary" nuclear 
matter in the outer layers of the star, for instance within a D3/D7 approach \cite{Hoyos:2016zke,Annala:2017tqz}, which can be improved by implementing a running  coupling \cite{Fadafa:2019euu,BitaghsirFadafan:2020otb}. "Bottom-up" models, inspired by but not rigorously based on a string theory, have also been employed, for example the so-called V-QCD model. It is valid in the Veneziano limit, which alleviates the large-$N_c$ artifacts, and has mostly been used for quark matter, but also allows for a description of nuclear matter and the high-density deconfinement phase transition.  In combination with known low-density approaches such as chiral effective field theory it has been used to model compact stars, including simulations of neutron star mergers  \cite{Jokela:2018ers,Ishii:2019gta,Ecker:2019xrw,Chesler:2019osn,Demircik:2020jkc,Jokela:2020piw}. 
For other holographic approaches to compact stars see for instance Refs.\ \cite{Ghoroku:2013gja,Ghoroku:2021fos}. 

In this paper, we employ the Witten-Sakai-Sugimoto model \cite{Witten:1998zw,Sakai:2004cn,Sakai:2005yt}, a "top-down" approach based on type-IIA string theory. We go beyond the previous holographic studies of neutron stars in the sense that our approach allows us to implement equilibrium with respect to the electroweak interaction and electric charge neutrality, two  essential conditions for realistic neutron stars that are routinely implemented in field-theoretical approaches.  We can therefore construct a mixed phase of our holographic baryonic matter with a gas of leptons, which models the crust of the star. Such a unified approach is highly desired even beyond the realm of holography \cite{Douchin:2001sv,Fortin:2016hny,Pais:2016xiu,Pearson:2020bxz,Rather:2020gja,Grams:2021lzx} and enables us to determine the location of the crust-core boundary fully dynamically.    

Baryons in the Witten-Sakai-Sugimoto model correspond to instantons of the gauge theory in the bulk \cite{Hata:2007mb,Hashimoto:2008zw}, and dense baryonic matter has been 
described in approximations of various degree of sophistication. The pioneering works 
used a pointlike approximation \cite{Bergman:2007wp} and -- in a complementary approach -- a homogeneous ansatz for the gauge fields \cite{Rozali:2007rx}, later refined in Ref.\ \cite{Elliot-Ripley:2016uwb}. Improvements to the pointlike approach included a multi-layered structure in the holographic direction \cite{Kaplunovsky:2012gb} and the construction of quarkyonic matter \cite{Kovensky:2020xif}.  Non-pointlike instantonic matter was considered in Refs.\ \cite{Li:2015uea,Preis:2016fsp}, including interactions of the instantons in the bulk \cite{BitaghsirFadafan:2018uzs}. 

Here we shall, firstly, consider the pointlike approximation. It was shown already in Ref.\ \cite{Zhang:2019tqd} that this approximation cannot account for realistic tidal deformabilities and masses at the same time. We confirm this conclusion and show that the energetically more favorable 2-layered solution \cite{Kaplunovsky:2012gb,Kovensky:2020xif} leads to similar results and does not produce realistic neutron stars
either. Secondly, and mainly, we then use the homogeneous ansatz of Ref.\ \cite{Rozali:2007rx} in the improved version of \cite{Kovensky:2021ddl}, where an isospin asymmetry 
for two-flavor baryonic matter was introduced. This improvement provides the setup to compute the thermodynamic properties of baryonic matter for arbitrary baryon and isospin chemical potentials. Thus, by adding a lepton gas we can construct (globally or locally) neutral dense matter in beta equilibrium. The calculation will be  performed  in the confined geometry of the model with antipodally separated flavor branes, which is the original version of the model introduced by Sakai and Sugimoto and which only has two free parameters, the 't Hooft coupling and the Kaluza-Klein scale. As a consequence of this simple setting,  our calculations will be independent of temperature. Also, we shall not include strangeness, quark matter, quarkyonic matter
or any form of Cooper pairing of the baryons. 

With the equation of state computed from the holographic model and with the help of the Tolman-Oppenheimer-Volkoff (TOV) equations we then compute masses, radii, and tidal deformabilities of the resulting stars and discuss their properties, most notably those of the holographic crust and the comparison of our results with the latest experimental constraints from gravitational waves \cite{LIGOScientific:2018hze,LIGOScientific:2020zkf}, the  mass measurement of the heaviest known neutron star \cite{NANOGrav:2019jur}, and the estimates for neutron star radii from the NICER mission \cite{Riley:2019yda,Miller:2019cac,Riley:2021pdl,Miller:2021qha}.

Our paper is organized as follows. Sec.\ \ref{sec:pointlike} is devoted to the pointlike approximation, including the holographic setup in Sec.\ \ref{sec:holsetuppointlike} and the evaluation and discussion of the results in Sec.\ \ref{sec:absence}.
Secs.\ \ref{sec:TOV} and \ref{sec:astroconstr} contain the TOV equations and collect the relevant astrophysical constraints, which are also needed for Sec.\ \ref{sec:beta}. Apart from these subsections, readers only interested in our more realistic approach may go directly to Sec.\ \ref{sec:beta}, where we discuss the neutron stars constructed from the homogeneous ansatz for baryonic matter. We briefly summarize the holographic setup in Sec.\ \ref{sec:holsetuphom} and discuss the addition of leptons and the construction of the crust in Secs.\ \ref{sec:leptons} and \ref{sec:crust}. The numerical results for the thermodynamics are presented in Sec.\ \ref{sec:thermo} and for the properties of the resulting stars in Sec.\ \ref{sec:NSs}. A summary and an outlook are given in Sec.\ \ref{sec:summary}.

\section{Pointlike baryons}
\label{sec:pointlike}

We start with the discussion of the simplest approximation of holographic baryonic matter, where
the instantons in the bulk are approximated by delta peaks and assumed to be non-interacting \cite{Bergman:2007wp} (this does not mean that the baryons on the field theory side are non-interacting). In this approximation, the number of flavors $N_f$ simply appears as a prefactor in the action, and thus the baryons can be thought of as objects composed of $N_c$ quarks of a single flavor.  Moreover, the baryon onset turns out to be of second order. In other words, matter at a nonzero baryon density always has nonzero positive pressure and thus cannot coexist with the vacuum. Therefore, baryonic matter in this approximation is obviously different from ordinary nuclear matter, which is made of neutrons and protons, and which, in the isospin symmetric case,
does have a nonzero saturation density with zero pressure. Despite these shortcomings it is a useful first step to start with this approximation. It allows us to connect our results to the previous literature, and it will be instructive to contrast our more realistic calculation in Sec.\ \ref{sec:beta} with the results from pointlike baryons.

\subsection{Holographic calculation}
\label{sec:holsetuppointlike}

The holographic setup needed for this section can be found in previous works \cite{Bergman:2007wp,Preis:2011sp,Li:2015uea,Kovensky:2020xif}. We shall therefore simply collect and only briefly explain the main equations needed to compute the equation of state, without going through the derivations or the details of the model. 

In this section, we work with the deconfined background geometry and allow for the D8- and $\overline{\rm D8}$-branes ("flavor branes") to have an arbitrary asymptotic separation $L$, such that together with the 't Hooft coupling $\lambda$ and the Kaluza-Klein mass $M_{\rm KK}$ the model has 3 free parameters. (We shall later set $N_f=2$ and $N_c=3$.) The flavor branes connect in the bulk at the point $u=u_c$, where $u$ is the coordinate of the holographic direction. This is interpreted as spontaneous breaking of chiral symmetry since the gauge theory on the flavor branes corresponds to the global flavor symmetry of the field theory on the boundary. We shall work with zero current quark masses throughout the paper for simplicity (a nonzero mass can be included following Refs.\ \cite{Kovensky:2019bih,Kovensky:2020xif}). In the following we may restrict our calculation without loss of generality to one half of the connected flavor branes, $u\in [u_c,\infty]$. 

The action, composed of Dirac-Born-Infeld and Chern-Simons terms, including the source term for the baryons, is
\bea \label{action1}
S &=& {\cal N} N_f \frac{V}{T}\int_{u_c}^\infty \hspace{-0.2cm}du\,\left\{u^{5/2}\sqrt{1+u^3f_T(u)x_4'^2(u)-\hat{a}_0'^2(u)} \right.\non[2ex]
&&\left.+\bar{n}_B\left[\frac{u}{3}\sqrt{f_T(u)}-\hat{a}_0(u)\right]\delta(u-u_B)\right\} \, , 
\eea
where prime denotes derivative with respect to $u$, $V$ is the spatial volume, $T$ is the temperature, and 
\be \label{N}
        {\cal{N}} \equiv \frac{N_c \MKK^4\lambda_0^3}{6\pi^2} \,, 
    \ee
with $\lambda_0 = \lambda/(4\pi)$, and where 
\be
f_T(u) = 1 - \frac{u_T^3}{u^3}   
\ee
is the blackening factor of the background metric, 
with the location of the horizon $u_T$, related to temperature via $4\pi T/M_{\rm KK} = 3u_T^{1/2}$.     
The action is a functional of the abelian part of the gauge field $\hat{a}_0(u)$ and the embedding function $x_4(u)$ of the flavor branes. The embedding gives the geometric shape of the branes in the $x_4$-$u$ subspace of the 10-dimensional background, where the $x_4$ direction is compactified with a radius given by $M_{\mathrm{KK}}^{-1}$. Pointlike baryons are placed at the point $u=u_B$, which, in principle, has to be determined dynamically. Due to the symmetry of the two halves of the branes, $u_B>u_c$ corresponds to a configuration of two layers of baryons. We shall also include the simplest case of a single baryon layer, where the baryons are forced to sit at the tip of the connected branes, $u_B=u_c$, as in the original work \cite{Bergman:2007wp}.  We have denoted the baryon number density, which is proportional to the number density of the delta-like instantons, by $\bar{n}_B$. Following Refs.\ \cite{Li:2015uea,Preis:2016fsp,BitaghsirFadafan:2018uzs,Kovensky:2019bih,Kovensky:2020xif}, the action is formulated in terms of dimensionless quantities (the only dimensionful quantities in Eq.\ (\ref{action1}) are $V$, $T$, and ${\cal N}$), and the physical, dimensionful baryon number density $n_B$ is
\be \label{nBbar}
n_B = \frac{N_f\lambda_0^2M_{\rm KK}^3}{6\pi^2}\bar{n}_B \, .
\ee
The equations of motion for $\hat{a}_0$ and $x_4$ can be solved algebraically for 
$\hat{a}_0'$ and $x_4'$,
\be
x_4' = \frac{k}{u^{11/2}f_T(u)}\zeta(u) \, , \quad \hat{a}_0'=\frac{\bar{n}_B\Theta(u-u_B)}{u^{5/2}}\,\zeta(u) \, ,
\ee
where $k$ is an integration constant, and 
\be
\zeta\equiv \left[1-\frac{k^2}{u^8f_T(u)}+\frac{\bar{n}_B^2\Theta(u-u_B)}{u^5}\right]^{-1/2} \, . 
\ee
The on-shell action (times $T/V$) is identified with the free energy density 
\be\label{NNf}
\Omega = {\cal N} N_f \bar{\Omega} \, ,
\ee
with the dimensionless version
\be \label{barOmega}
\bar{\Omega} = \int_{u_c}^\infty du\,u^{5/2} [\zeta(u)-1] - \frac{2}{7} u_c^{7/2} +p_0 \, .
\ee
Here we have used that stationarity of the on-shell action with respect to 
$\bar{n}_B$ yields 
\be\label{a0uB}
\hat{a}_0(u_B) = \frac{u_B}{3}\sqrt{f_T(u_B)} \, ,
\ee
and we have subtracted the infinite, medium-independent vacuum contribution $\frac{2}{7}\Lambda^{7/2}-p_0$, where $\Lambda$ is an ultraviolet cutoff 
(which is sent to infinity after the subtraction). The finite contribution 
\be
p_0 = \frac{2^{15}\pi^4}{7 \ell^7}\left(\frac{\Gamma[9/16]}{\Gamma[1/16]}\right)^8  
\ee
has been included in the subtraction to normalize the vacuum pressure to 0, i.e., $\bar{\Omega}=0$ for $u_T=\bar{n}_B=0$. Here, $\ell=M_{\rm KK} L$ is the dimensionless version of the asymptotic separation of the flavor branes, and we have used that in the vacuum $k=u_c^4$ and 
\be \label{uc0}
u_c = \frac{16\pi}{\ell^2} \left(\frac{\Gamma[9/16]}{\Gamma[1/16]}\right)^2  \, .
\ee

 The dimensionless (quark) chemical potential is introduced as the boundary value of the abelian part of the gauge field, $\bar{\mu}_B = \hat{a}_0(\infty)$, and is related to its dimensionful (baryon) counterpart by
\be \label{muBmu}
\mu_B = N_c \lambda_0M_{\rm KK} \bar{\mu}_B \, .
\ee
Using Eq.\ (\ref{a0uB}), we can thus write the boundary conditions for $x_4$ and $\hat{a}_0$ as 
\bea
\frac{\ell}{2} &=& \int_{u_c}^\infty du\, x_4' \, , \quad 
\bar{\mu}_B = \frac{u_B}{3}\sqrt{f_T(u_B)} +\int_{u_B}^\infty du\,\hat{a}_0' 
 \, . \hspace{0.6cm} \label{ellmu}
\eea
For the minimization with respect to $u_c$ one has to treat the 1-layer and 2-layer cases separately. In both cases, this condition can be solved explicitly for $k$, 
\bea 
k^2&=&(u_c^8+\eta u_c^3\bar{n}_B^2)f_T(u_c) \non[2ex]
&&-\eta u_c^3\left(\frac{\bar{n}_B}{3}\right)^2\left[\frac{3-f_T(u_c)}{2}\right]^2 \, ,  \label{k12}
\eea
where $\eta=1$ ($\eta=0$) for 1 layer (2 layers)\footnote{For the 2-layer case see Ref.\ \cite{Kovensky:2020xif}, and for the 1-layer case see for instance appendix B of Ref.\ \cite{Li:2015uea}, where Eqs.\ (B7) and (B8) contain a typo: $1+f_T(u_c)$ has to be replaced by $3-f_T(u_c)$, as in our Eq.\ (\ref{k12}).}. In the 2-layer case, the additional condition of minimizing the free energy density with respect to $u_B$ yields 
\bea 
\bar{n}_B &=& \frac{6\Delta_T(u_B)}{u_B^{3/2}} \sqrt{\frac{
u_B^8 f_T(u_B) - k^2 }{
f_T(u_B) [9-\Delta_T^2(u_B)]^2 }}\, , \label{knB2}
\eea
where we have abbreviated
\begin{equation}
    \Delta_T(u_B)  \equiv  
    \frac{1}{\sqrt{f_T(u_B)}}\left(
    1+ \frac{u_T^3}{2u_B^3}
    \right)
    \, .
\end{equation}
One  now proceeds by solving Eqs.\ (\ref{ellmu}) with the help of Eqs.\ (\ref{k12}) and (\ref{knB2}). 
The asymptotic separation $\ell$ can be eliminated  by working with the variables  $\tilde{n}_B = \bar{n}_B\ell^5$, $\tilde{\mu}_B=\bar{\mu}_B\ell^2$, $\tilde{u}_c=u_c\ell^2$, $\tilde{u}_B=u_B\ell^2$, which yields the free energy density $\tilde{\Omega}=\bar{\Omega}\ell^7$. 
Since our choice of the dimensionless variables already absorbs all other model parameters, Eqs.\ (\ref{ellmu}) only have to be solved once, without fixing $\lambda$, $M_{\rm KK}$, and $L$, which only become relevant in the translation to 
physical results. 
Having in mind the "decompactified limit" of the model $\ell \ll 1$, where the confined phase is ignored \cite{Preis:2011sp,Kovensky:2020xif},  we shall restrict ourselves to zero temperature. In this case, the second-order baryon onset occurs at
\be
\bar{\mu}_B=\frac{u_c}{3} \, ,
\ee
with $u_c$ from Eq.\ (\ref{uc0}). 
This is true for both 1-layer and 2-layer configurations. In the 2-layer case the baryons sit at $u_B=u_c$ at the onset and move away from this point towards the ultraviolet as the chemical potential (and thus the baryon density) is increased. The free energies of the two configurations are identical only at the onset; as soon as the chemical potential increases, the 2-layer solution is energetically preferred. Multi-layer solutions are expected to take over at large densities \cite{Kaplunovsky:2012gb,Kovensky:2020xif}, and they should be taken into account in principle. However, we shall demonstrate that the 1-layer and 2-layer solutions do not differ much in their effect on compact star properties. And, since we shall see that realistic constraints are far from being met within the present approximation, we do not expect multi-layer solutions to change this conclusion.  Also, although the pointlike approximation suggests that multi-layer solutions become preferred at large densities, calculations allowing for finite-width instantons indicate that more than 2 layers are never preferred \cite{Preis:2016fsp,Elliot-Ripley:2016uwb}. We therefore ignore configurations with more than two baryon layers. 

\begin{figure} [t]
\begin{center}
\includegraphics[width=\columnwidth]{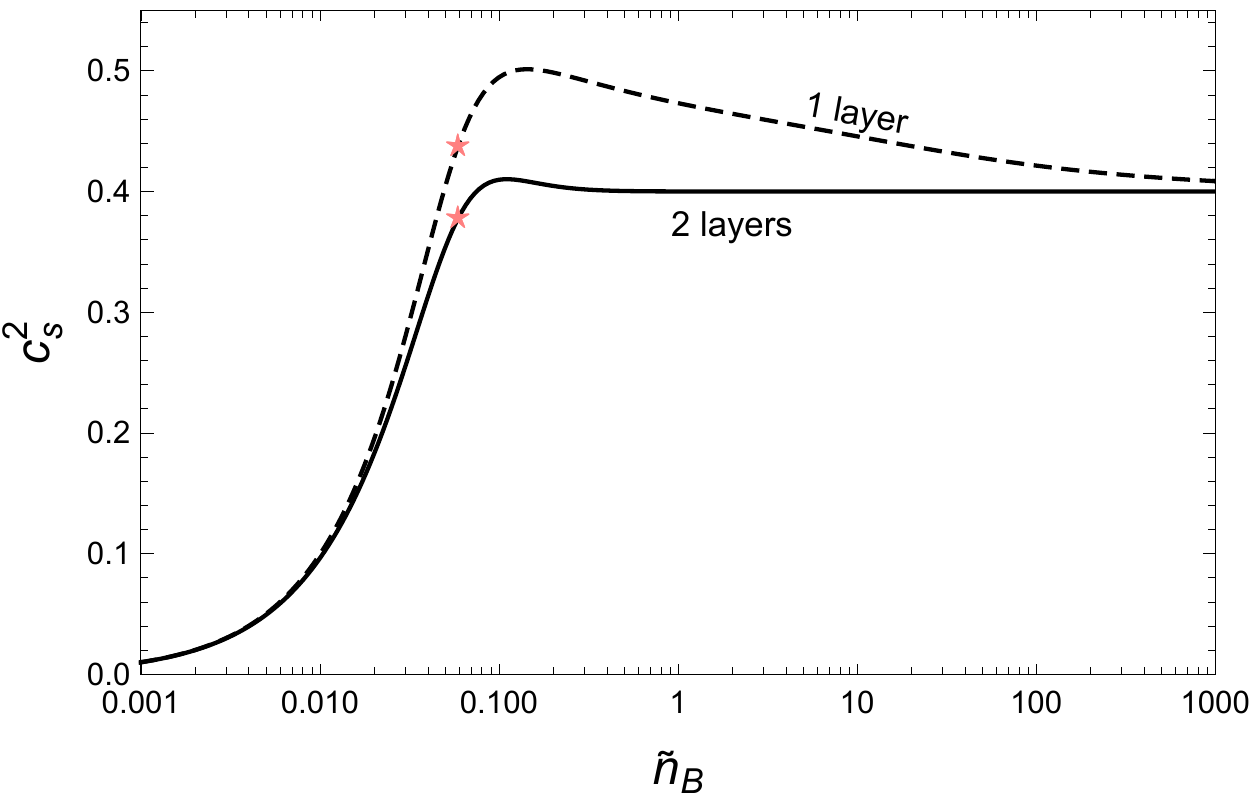}
\caption{Speed of sound squared as a function of the dimensionless baryon number density $\tilde{n}_B=\bar{n}_B\ell^5$ for 1-layer (dashed) and 2-layer (solid) pointlike baryonic matter. The  stars indicate the density and speed of sound in the center of the most massive stars. 
}
\label{fig:sound}
\end{center}
\end{figure}

To obtain the equation of state, the solutions of Eqs.\ (\ref{ellmu}) 
are inserted back into the free energy density
(\ref{barOmega}), which yields the pressure $\bar{P} = -\bar{\Omega}$ and the zero-temperature energy density 
\be
\bar{\epsilon} = \bar{\Omega} +\bar{\mu}_B\bar{n}_B \, .
\ee
Dimensionless pressure and energy density are both translated to their dimensionful counterparts $P$ and $\epsilon$ with the factor given in Eq.\ (\ref{NNf}).
Finally, we will also need the speed of sound 
\be
c_s^2 = \frac{\partial P}{\partial \epsilon} = \frac{n_B}{\mu_B}\left(\frac{\partial n_B}{\partial\mu_B}\right)^{-1} \, .
\ee
Here, the derivative with respect to $\epsilon$ is in general taken at fixed entropy per particle, and the right-hand side is valid at zero temperature. We show the results for the 1-layer and 2-layer configurations in Fig.\ \ref{fig:sound}. In both cases the speed of sound is non-monotonic as a function of density and approaches the value $c_s^2=2/5$ for asymptotically large densities (in an asymptotically free theory such as QCD the result approaches 1/3). Interestingly, in the 2-layer configuration the speed of sound approaches the asymptotic value much faster. We have indicated the value of the central density within the most massive  compact star -- to be computed in the following subsections -- for both cases, which 
shows that the non-monotonicity plays no role in the interior of any star. 

\subsection{TOV equations and tidal deformability}
\label{sec:TOV}

We now combine the thermodynamics with gravity to construct stars in hydrostatic equilibrium. To this end, we employ the well-known TOV equations \cite{tolman,PhysRev.55.364,PhysRev.55.374},  restricting ourselves to the static, spherically symmetric case,
\begin{subequations} \label{TOVs}
\bea
&&\frac{\partial P}{\partial r} = -\frac{G}{r^2} \frac{(M+4\pi Pr^3)(\epsilon+P)}{1-\frac{2GM}{r}} \, , \label{TOV1}\\[2ex]
&& \frac{\partial M}{\partial r} = 4\pi r^2 \epsilon \, , \label{TOV2}\\[2ex]
&&0= r\frac{\partial y}{\partial r}+y^2+\frac{4\pi G r^2\left(5\epsilon+9P+\frac{\epsilon+P}{c_s^2}\right)-6}{1-\frac{2GM}{r}}\non[2ex]
&&+\frac{y\left[1-4\pi G r^2(\epsilon-P)\right]}{1-\frac{2GM}{r}}-\frac{4G^2(M+4\pi Pr^3)^2}{r^2\left(1-\frac{2GM}{r}\right)^{2}} \, , \hspace{0.5cm}\label{TOV3}
\eea
\end{subequations}
where $G=6.709 \times 10^{-39}\, {\rm GeV}^{-2}$ is the gravitational constant.
The third equation (\ref{TOV3}) is added to the TOV equations (\ref{TOV1}) and (\ref{TOV2}) to compute the tidal deformability \cite{1967ApJ...149..591T,Hinderer:2007mb,Damour:2009zoi,Postnikov:2010yn}. It contains the function $y(r)$, which is related to the metric perturbation from tidal deformations. The equations are solved for mass $M(r)$, pressure $P(r)$, and $y(r)$ as functions of the radial coordinate $r$ after employing an equation of state $\epsilon(P)$ and using the boundary conditions in the center of the star $M(0)=0$, $P(0)=P_c$, $y(0)=2$, where $P_c$ is the central pressure. Varying $P_c$ yields all possible stars for a given equation of state. Then, $P(R)=0$ (the vacuum pressure being set to zero) defines the radius of the star $R$, and the total gravitational mass of the star is $M(R)$, which we shall simply denote by $M$ in the following.  
The tidal deformability 
\be
\Lambda = \frac{2k_2}{3c^5} 
\ee
is then computed from the compactness of the star, 
\be \label{comp}
c= \frac{GM}{R} \, ,
\ee
and the so-called tidal Love number \cite{1909RSPSA..82...73L}
\bea
&&k_2 = \frac{8c^5}{5}(1-2c)^2[2-y_R+2c(y_R-1)]\non[2ex]
&&\times \Big\{2c[6-3y_R+3c(5y_R-8)]\non[2ex]
&&+4c^3[13-11y_R+c(3y_R-2)+2c^2(1+y_R)]\non[2ex]
&& +3(1-2c)^2[2-y_R+2c(y_R-1)]\ln(1-2c)\Big\}^{-1} \, , \hspace{1cm}
\eea
where $y_R\equiv y(R)$.
We shall perform the calculation with dimensionless quantities 
\be \label{tilde}
\hat{r} = \frac{r}{r_0} \, , \quad \hat{M} = \frac{M}{M_0} \, , \quad \hat{P}=\frac{P}{\epsilon_0} \, , \quad 
\hat{\epsilon} = \frac{\epsilon}{\epsilon_0} \, , 
\ee
where we choose the scales $r_0$, $M_0$, $\epsilon_0$ to obey 
\be \label{AB1}
1 = \frac{GM_0}{r_0} = \frac{4\pi r_0^3\epsilon_0}{M_0} \, .
\ee
The benefit of this choice is that the dimensionless version of Eqs.\ (\ref{TOVs}) does not explicitly depend on  $r_0,M_0,\epsilon_0$, which only become relevant to transform back to the dimensionful results. 
Since Eq.\ (\ref{AB1}) leaves one freedom for the three scales, we will choose $\epsilon_0$ to our convenience according to the holographic calculation and then determine $M_0$ and $r_0$ from that choice.

\subsection{Astrophysical constraints}
\label{sec:astroconstr}

Here and in the rest of the paper we shall have in mind the following constraints from astrophysical data: 
\begin{itemize}

\item Each mass-radius curve must allow for a mass of about 2.1 solar masses according to the heaviest currently  known neutron star \cite{NANOGrav:2019jur}; if the compact object from the merger GW190814 with a black hole is a neutron star, the lower limit for the maximal mass is even 2.5 solar masses or higher \cite{LIGOScientific:2020zkf}.

\item The tidal deformability for a roughly 1.4 solar mass star was constrained by the merger of two neutron stars GW170817 to be about $70 \lesssim \Lambda_{1.4} \lesssim 580$ \cite{LIGOScientific:2018hze}.

\item Data from the NICER collaboration was used to estimate the radius of the above mentioned heaviest neutron star to be about $(11.4 - 13.7)\,{\rm km}$ \cite{Riley:2021pdl} and $(12.2 - 16.3)\,{\rm km}$ \cite{Miller:2021qha}, while previous data for a roughly 1.4 solar mass star was used to obtain radius estimates of about $(11.5 - 13.9)\,{\rm km}$ \cite{Riley:2019yda} and $(12.0 - 14.3)\,{\rm km}$ \cite{Miller:2019cac}. 

\end{itemize}
\subsection{Absence of realistic stars with pointlike baryons}
\label{sec:absence}

In the case of pointlike baryons, we compute the free energy density $\tilde{\Omega}=\bar{\Omega}\ell^7$, with $\bar{\Omega}$ defined in Eq.\ (\ref{NNf}). Therefore,  the obvious choice for the scale $\epsilon_0$ is
\be \label{eps0pointlike}
\epsilon_0 = \frac{N_f{\cal N}}{\ell^7} \simeq 6.214\times 10^4\left(\frac{K}{\rm GeV}\right)^4\frac{\rm MeV}{{\rm fm}^3} \, ,
\ee
where we have 
introduced the energy scale 
\be\label{Kdef}
K\equiv \left(\frac{N_fN_c\lambda_0^3}{9\pi L^7M_{\rm KK}^3}\right)^{1/4} \, . 
\ee
With Eq.\ (\ref{AB1}) we then find
\begin{subequations}
\bea
M_0&\simeq& 0.666 \left(\frac{K}{{\rm GeV}}\right)^{-2} M_\odot \, , \\[2ex]
r_0&\simeq& 0.984 \left(\frac{K}{{\rm GeV}}\right)^{-2} {\rm km} \, ,
\eea
\end{subequations}
where $M_\odot=1.988\times 10^{30}\, {\rm kg}$ denotes the mass of the sun. 
We observe that mass, radius, and tidal deformability of the star depend solely on the particular combination of the model parameters in Eq.\ (\ref{Kdef}).

\begin{figure} [t]
\begin{center}
\includegraphics[width=\columnwidth]{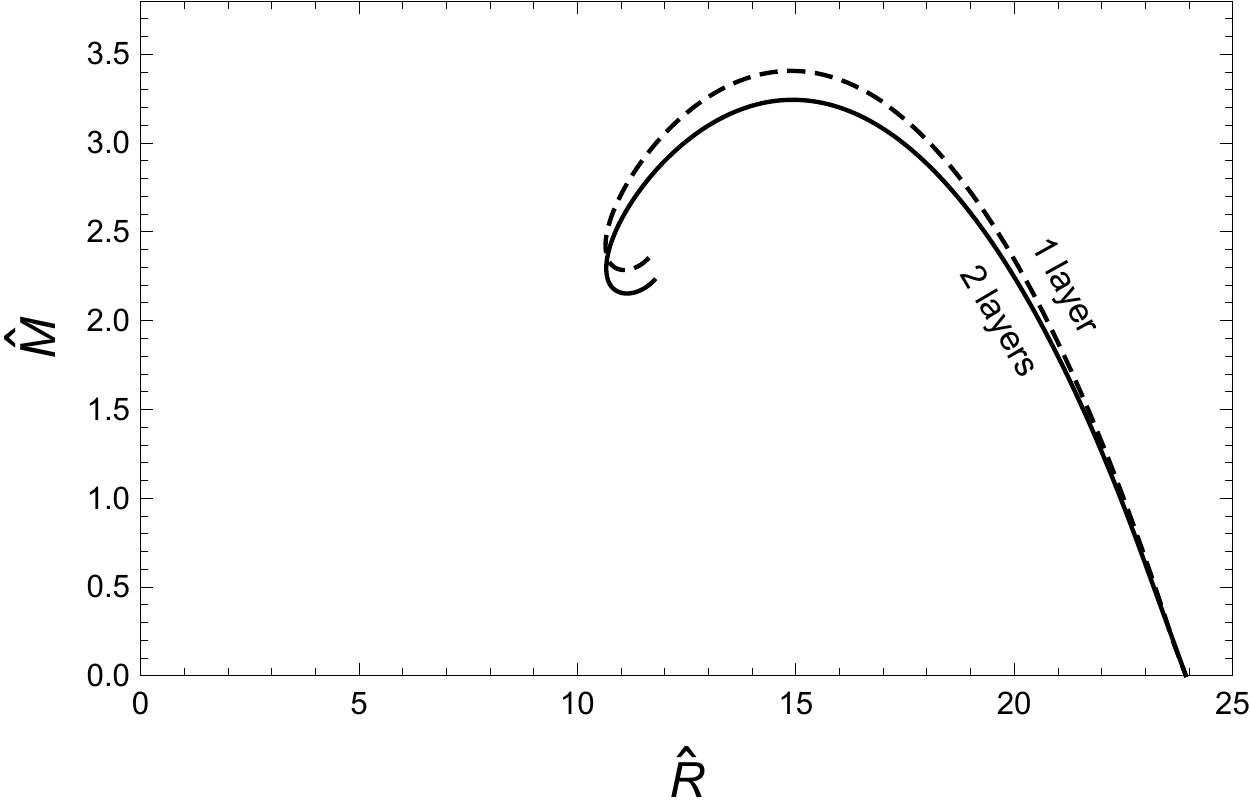}
\includegraphics[width=\columnwidth]{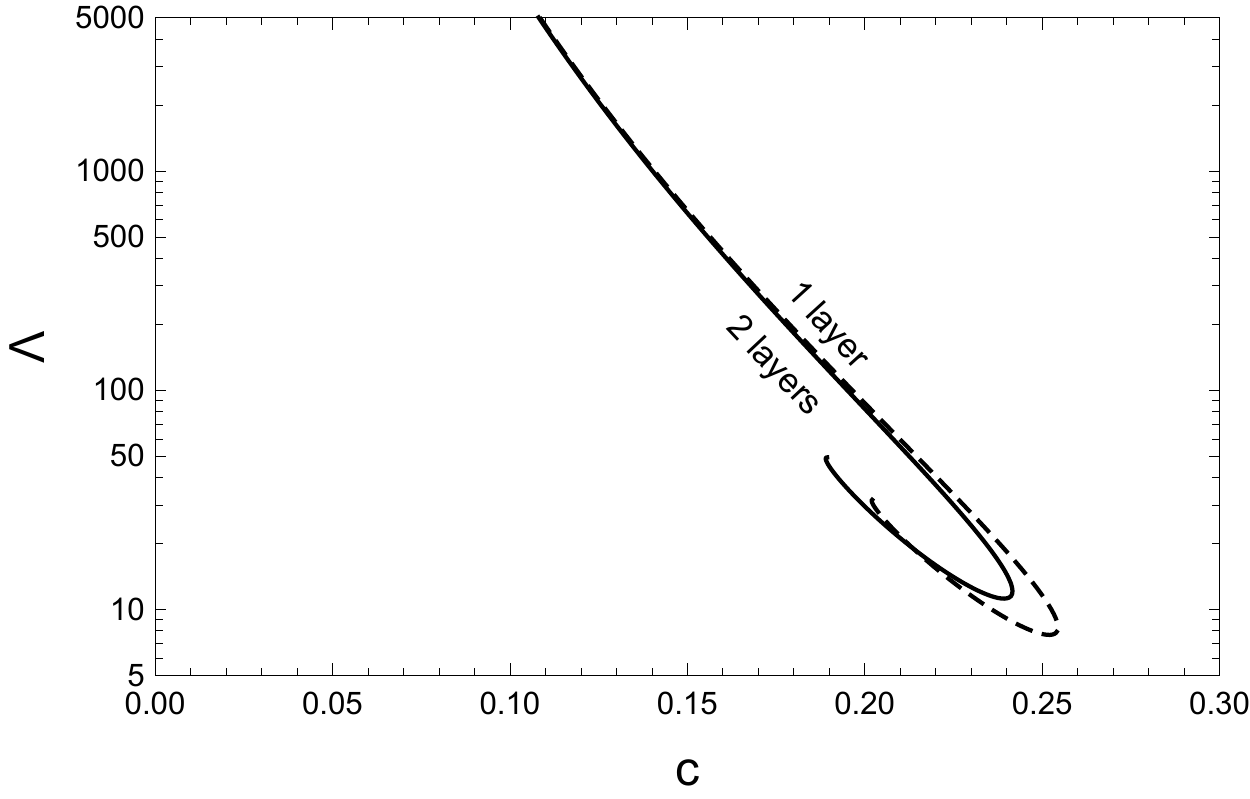}
\caption{Mass-radius curve in dimensionless units (upper panel) and tidal deformability as a function of compactness (lower panel) for 1-layer (dashed) and 2-layer (solid) configurations of pointlike baryons. Both plots are independent of the model parameters.}
\label{fig:MRdimless}
\end{center}
\end{figure}

\begin{figure} [t]
\begin{center}
\includegraphics[width=\columnwidth]{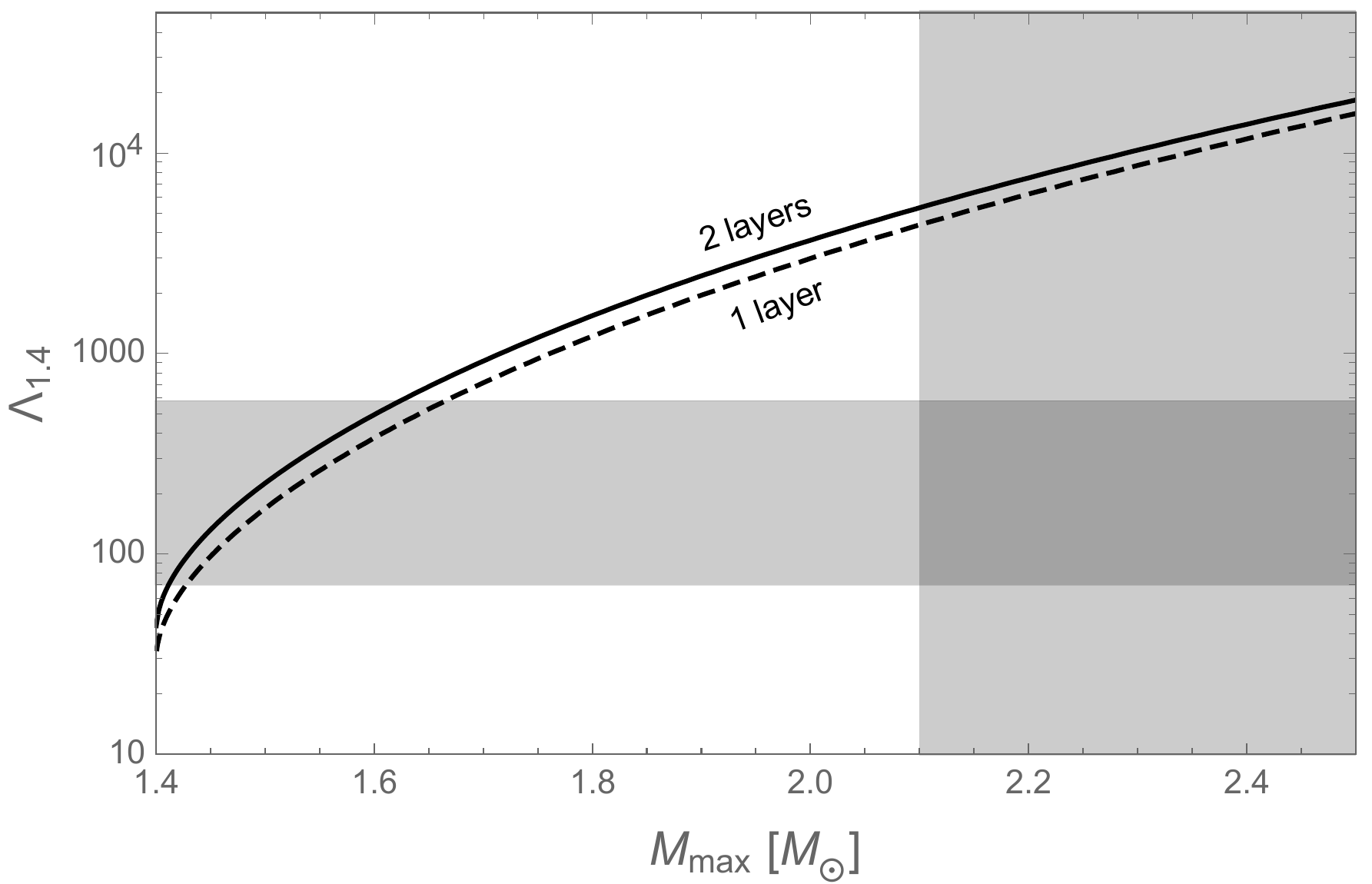}
\caption{Relation between maximal masses and tidal deformabilities for 1.4 solar mass stars using the pointlike approximation for baryons. The curve is parameterized 
by the energy scale $K$, containing the only relevant combination of the three model parameters, from small $K$ at the upper right to large $K$ to the lower left; if $K\gtrsim 1.24\, {\rm GeV}$ ($1.27 \, {\rm GeV}$) for 2 layers (1 layer)  all stars have masses below 1.4 solar masses. The grey bands indicate the astrophysical constraints, see 
Sec.\ \ref{sec:astroconstr}, and we see that in the pointlike approximation no choice of the model parameters can meet them simultaneously. 
}
\label{fig:LamMmax}
\end{center}
\end{figure}

Solving the coupled equations (\ref{TOVs}) in their dimensionless version numerically for different central pressures yields the mass-radius curves shown in Fig.\ \ref{fig:MRdimless}. We see that 1-layer and 2-layer phases yield very similar results, with the 2-layer solution allowing for a slightly smaller maximal mass. In the plot we have extended the curves to radii smaller than that of the maximum mass star, although this part of the curve corresponds to unstable stars with respect to radial oscillations \cite{1966ApJ...145..505B,glendenning}. As the central pressure is decreased, the radius of the stars asymptotes to a finite value of about $\hat{R}\simeq 23.9$. To understand this we observe (numerically) that for small densities the equation of state behaves like $P\propto \epsilon^2$. In this regime the Lane-Emden equation is applicable, obtained in the Newtonian limit and with a polytropic equation of state from the TOV equations, see for instance Ref.\ \cite{Schaffner-Bielich:2020psc}. From this equation one can check analytically that a quadratic equation of state results in a fixed radius independent of the mass. Figure \ref{fig:MRdimless} also includes the tidal deformability as a function of compactness. For both plots no choice of the model parameters was necessary. To connect our results to 
the astrophysical data we need to discuss their dependence on the energy scale $K$.

As a first choice, let us fit our model parameters to basic QCD vacuum properties. For instance we may use the pion decay constant $f_\pi \simeq 93\, {\rm MeV}$ and interpret the value of the chemical potential at the baryon onset as the vacuum mass of the nucleon $m_N\simeq 939\, {\rm MeV}$. Using the expression for the pion decay constant for the deconfined geometry of the Witten-Sakai-Sugimoto model \cite{Kovensky:2020xif,Callebaut:2011ab}, reinstating the dimensionful factor for the chemical potential (\ref{muBmu}) and using Eq.\ (\ref{uc0}), this yields 
\begin{subequations}\label{mNfpi}
\bea
f_\pi^2 &=& \frac{32 (\lambda/\ell) N_c }{3\pi^2 L^2} \left(\frac{\Gamma[9/16]}{\Gamma[1/16]}\right)^3\frac{\Gamma[11/16]}{\Gamma[3/16]}
\, , \\[2ex]
m_N &=& \frac{4(\lambda/\ell)N_c}{3 L} \left(\frac{\Gamma[9/16]}{\Gamma[1/16]}\right)^2
 \, , 
\eea
\end{subequations}
from which we obtain $K\simeq0.842\, {\rm GeV}$. Here we have set $N_c=3$ (extrapolating down from large $N_c$) and $N_f=2$. 
We find a maximal mass $M_{\rm max} \simeq 3.0\, M_\odot$
($3.2 \,M_\odot$) for 2 layers (1 layer), a tidal deformability of a 1.4 solar mass star $\Lambda_{1.4}\simeq 6.6\times 10^4$ ($8.2 \times 10^4$) and a corresponding radius $R_{1.4} \simeq 30 \, {\rm km}$ $(31\, {\rm km})$. The results for $\Lambda_{1.4}$ and $R_{1.4}$ are far beyond the constraints from astrophysical data, and thus we do not even come close to reproducing vacuum and neutron star properties at the same time within the pointlike approximation. 

We may lower our expectations and ask if there is any parameter choice that at least fulfills the astrophysical constraints, ignoring $m_N$ and $f_\pi$. Since the translation of the dimensionless results into physical units only 
requires the choice of the single scale $K$, it is easy to explore the
entire parameter space. Each $K$ gives a maximal mass $M_{\rm max}$ and a tidal deformability $\Lambda_{1.4}$.
By varying $K$ we can thus construct the curves shown in Fig.\ \ref{fig:LamMmax} for 1-layer and 2-layer configurations. The shaded regions show the constraints from the data, and we see that our curve does not enter the area were both constraints are fulfilled simultaneously. For instance, the largest possible mass compatible with the realistic band for the tidal deformability is about $1.62 \, M_\odot$ ($1.67\,M_{\odot}$) for 2 layers (1 layer), in accordance with the 
conclusion for 1 layer in Ref.\ \cite{Zhang:2019tqd}.  
Since the curves capture all possibilities, we can rigorously conclude that the present approximation, while being useful as a first attempt due to its simplicity, cannot produce  realistic neutron stars. We therefore move on to a different approximation of holographic baryons which shall turn out to be more relevant for astrophysical applications.

\section{Beta-equilibrated, electrically neutral, holographic neutron stars}
\label{sec:beta}

We have seen that the pointlike approximation of the previous section can only be a first step at best towards a holographic description of dense matter inside neutron stars. More sophisticated approaches in the Witten-Sakai-Sugimoto model based on instantons with a nonzero, dynamically determined, width do exist in the literature. However, 
so far none of these instanton-based approaches have included isospin, except for studies of single baryons \cite{Hata:2007mb}. Since an isospin asymmetry is a crucial ingredient for neutron star matter we will use a recently developed approach which is not explicitly built on instanton solutions but which does account for isospin \cite{Kovensky:2021ddl}. This approach makes use of an ansatz for the non-abelian part of the gauge fields that is homogeneous in position space \cite{Rozali:2007rx,Li:2015uea,Elliot-Ripley:2016uwb}. 

Here, we shall only use the simplest version of the formalism developed in Ref.\ \cite{Kovensky:2021ddl}. Namely, we work in the confined geometry, with antipodal separation of the flavor branes, and employ the Yang-Mills approximation for the action.
Moreover, we ignore the possibility of a pion condensate, which has been shown to coexist with baryonic matter in large parts of the phase diagram in the presence of an isospin chemical potential \cite{Kovensky:2021ddl}. This is an interesting observation also for applications to compact stars since it is still an open question whether pions form a condensate in neutron star matter  \cite{Migdal:1973zm,PhysRevLett.29.382,Akmal:1997ft}. However, in the approach of Ref.\ \cite{Kovensky:2021ddl} the holographic pion condensate cannot easily be separated from the baryonic contribution, and thus it is not straightforward to couple the system to electromagnetism, i.e., assign the correct electric charges to the nucleons and pions separately. Therefore 
(and since Ref.\ \cite{Kovensky:2021ddl} only provides the necessary equations in the limit of massless pions), we leave holographic pion condensation in neutron stars for future studies. 

Also, our restriction to the confined geometry of the Witten-Sakai-Sugimoto model is done for simplicity. The deconfined geometry -- used in the previous section for the pointlike baryons  -- has the advantage to be easily generalizable to nonzero temperatures and to allow for chirally symmetric phases. It may thus be used in principle to study hybrid stars with a quark matter or quarkyonic core.  (However, the transition densities to chirally symmetric matter appear to be very large, perhaps too large for neutron star interiors, at least in the approximations currently available in the literature.) For our purposes,  the deconfined geometry 
leads to numerically more challenging equations, which make a systematic study very cumbersome. We therefore work with the simpler equations from the confined background, where in particular the antipodal separation of the flavor branes simplifies the problem because their embedding does not have to be calculated dynamically. 

Moreover, in this section we do not include the possibility of more than one baryon layer in the holographic direction. As for the pointlike case of Sec.\ \ref{sec:pointlike}, this possibility should, strictly speaking, also be taken into account in the approach of the homogeneous ansatz. It was indeed studied in Ref.\ \cite{Elliot-Ripley:2016uwb}, albeit within a slightly different homogeneous ansatz, made on the level of the field strengths, not the gauge fields. A generalization of this ansatz  to nonzero isospin density, however, has not yet been performed. We recall from Sec.\ \ref{sec:pointlike} that within the pointlike approximation the results for 2 layers are very similar
to the 1-layer case. Therefore, although a full calculation has to ultimately show the quantitative effect of additional baryon layers, we do not expect them to alter our main conclusions significantly. 

Before collecting the relevant equations, let us point out one important shortcoming of the approach used here. Since it does not employ any quantization in the bulk, which can account for finite-$N_c$ effects and results in  discrete states for neutron and proton \cite{Hata:2007mb}, the baryon spectrum in the approximation used here is continuous in the isospin direction \cite{Kovensky:2021ddl}. As a consequence, symmetric nuclear matter can be thought of as being composed of predominantly isospin-symmetric baryons, and an isospin asymmetry in the many-body system requires the population of heavier, isospin-asymmetric states. This is in contrast to the real world, where isospin-symmetric nuclear matter is already made of baryons with nonzero isospin number, neutrons and protons, and an isospin asymmetry can be created by changing their population, without creating new baryons with different masses. An important consequence is that our holographic nuclear matter has an unrealistically large symmetry energy \cite{Kovensky:2021ddl} -- of the order of the baryon mass rather than about $30\, {\rm MeV}$ --  and we shall see that as a consequence our neutron star matter will have a very large proton fraction, much closer to symmetric matter than to pure neutron matter. Nevertheless, we shall see that qualitatively, and in many aspects also quantitatively, our approach yields realistic properties of neutron stars.

\subsection{Holographic setup}
\label{sec:holsetuphom}

In this subsection, we collect and briefly comment on the relevant equations of the holographic setup. More details and derivations can be found in Ref.\ \cite{Kovensky:2021ddl}. The action in the Yang-Mills approximation is 
\bea
    S &=& 
{\cal{N}} N_f \frac{V}{T} \int_{u_\mathrm{KK}}^\infty 
    \left[\frac{u^{5/2}}{2\sqrt{f}} \left(g_1 - f \hat{a}_0'^2 - f a_0'^2 + g_2 - g_3\right) \right.\non[2ex]
    &&\left.- \frac{9}{4} \lambda_0  \hat{a}_0  h^2 h'\right]\, du \, ,
    \label{Sconf}
\eea
with the abbreviations
\begin{equation}
    \label{g1g2g3}
g_1\equiv \frac{3fh'^2}{4} \, , \qquad g_2\equiv \frac{3\lambda_0^2h^4}{4u^3} \, , \qquad g_3\equiv \frac{2\lambda_0^2h^2a_0^2}{u^3} \, .
\end{equation}
As in Sec.\ \ref{sec:pointlike}, prime denotes derivative with respect to $u$. The structure of the action is similar to the one used for the pointlike baryons in the deconfined geometry (\ref{action1}). Besides the absence of the square root due to the Yang-Mills expansion, the different background geometry leads to a different, temperature-independent, metric factor 
\be
    f = 1-
    \frac{u_{\mathrm{KK}}^3}{u^3} \, ,
\ee
where $u_{\rm KK}$ is the location of the tip of the cigar-shaped $u$-$x_4$ subspace, i.e., 
$u\in [u_{\rm KK},\infty]$, again working on one half of the connected flavor branes. In our units, $u_{\rm KK}=4/9$. The action depends on the temporal component of the abelian $U(1)$ gauge field $\hat{a}_0$ with boundary condition $\bar{\mu}_B=\hat{a}_0(\infty)$ and the temporal component of the non-abelian $SU(2)$ part $a_0^a\sigma_a$, where $\sigma_a$ are the Pauli matrices. The isospin chemical potential acts as a boundary value for $a_0\equiv a_0^3$, $\bar{\mu}_I=a_0(\infty)$, and it is consistent to set $a_0^1=a_0^2=0$, at least in the chiral limit employed here. The spatial components of the non-abelian part are written as $a_i=-\lambda_0h\sigma_i/2$, with a single function $h(u)$ that generates nonzero baryon number through a discontinuity at $u=u_{\rm KK}$. We work with the same dimensionless units as in Sec.\ \ref{sec:pointlike}, such that the physical chemical potentials are obtained via Eq.\ (\ref{muBmu}). 

The equation of motion for $\hat{a}_0$ can easily be integrated to give 
\be
    \hat{a}_0' = \frac{\bar{n}_BQ}{u^{5/2}\sqrt{f}} \, ,
    \label{a0p}
\ee
where $Q(u)\equiv 1-h^3(u)/h_c^3$ with the boundary value $h_c \equiv h(u_{\rm KK}) = -[4\bar{n}_B/(3\lambda_0)]^{1/3}$.  The dimensionless baryon density $\bar{n}_B$ and the isospin density $\bar{n}_I$ (given below) are related to their dimensionful versions via Eq.\ (\ref{nBbar}). 
The equations of motion for 
$h$ and $a_0$ are 
\begin{subequations}
\label{EOMs}
\begin{eqnarray}
\left(
u^{5/2} \sqrt{f} a_0'
\right)'&=& \frac{2\lambda_0^2h^2a_0}{u^{1/2}\sqrt{f}} \, , \label{eomK} \\[2ex]
\frac{3}{2}
\left(u^{5/2}\sqrt{f}h'
\right)'-\frac{9\lambda_0h^2\bar{n}_BQ}{2u^{5/2}\sqrt{f}} &=& \frac{\lambda_0^2h(3h^2-4a_0^2)}{u^{1/2}\sqrt{f}} \, . \hspace{0.8cm}\label{eomh} 
\end{eqnarray}
\end{subequations}
These equations have to be solved numerically for $a_0(u)$ and $h(u)$ with the boundary conditions already given above, together with $h(\infty)=0$ and $a_0'(u_{\rm KK}) = 0$.
From the solutions we then compute baryon chemical potential and isospin number density
\begin{subequations} \label{muBnI}
\bea
    \bar{\mu}_B &=&  \frac{u_{\rm KK}^2 h_{(1)}}{2 \sqrt{3} \lambda_0 h_c^2} + \int_{u_{\rm KK}}^{\infty} du \, 
    \frac{\bar{n}_BQ}{u^{5/2}
    \sqrt{f}} \, , \label{muBnI1}\\[2ex]
    \bar{n}_I &=&   2\lambda_0^2 \int_{u_{\rm KK}}^\infty du\, 
   \frac{h^2 a_0}{u^{1/2}\sqrt{f}}  \, ,   \label{muBnI2}
\eea
\end{subequations}
where $h_{(1)}$ is numerically extracted from the behavior close to the tip of the connected branes, $h(u)= h_c +h_{(1)}\sqrt{u-u_{\rm KK}} +\ldots$. Inserting the solutions back into the action yields the (dimensionless) free energy density
\bea
   \bar{\Omega}_B&=& \int_{u_{\rm KK}}^\infty du\,\frac{u^{5/2}}{2\sqrt{f}}\left[g_1 + g_2+\frac{(\bar{n}_BQ)^2}{u^5}+\frac{2\lambda_0^2\bar{\mu}_Ih^2a_0}{u^3}\right]\non[2ex]
   &&-\bar{\mu}_B\bar{n}_B-
   \bar{\mu}_I \bar{n}_I \, .
   \label{solom}
\eea
The divergent part of the action is absent from the beginning due to the use of the 
Yang-Mills approximation, and the vacuum pressure is automatically normalized to zero, $\bar{\Omega}_B(\bar{n}_B=\bar{n}_I=0)=0$, i.e., no further vacuum subtraction is necessary. We have added a subscript $B$ to the free energy density to indicate that this is the baryonic part, to which the leptonic part is added now. 

\subsection{Adding leptons}
\label{sec:leptons}

In order to account for realistic neutron star matter we need to add leptons, which will serve to neutralize the system. The total (dimensionless) free energy density 
is thus the sum of baryonic and leptonic contributions, 
\be \label{barP}
\bar{\Omega}  =  \bar{\Omega}_B + \bar{\Omega}_\ell 
\, ,
\ee
where 
\be \label{pell}
\bar{\Omega}_\ell = \bar{\Omega}(m_e,\mu_e) + \bar{\Omega}(m_\mu,\mu_\mu) 
\ee
is the free energy density of non-interacting electrons and muons with masses $m_e=511\, {\rm keV}$, $m_\mu=106\, {\rm MeV}$, with the zero-temperature Fermi gas expression
\bea
\bar{\Omega}(m,\mu) &\equiv& -\frac{\Theta(\mu-m)}{24\pi^2N_f{\cal N}}\Bigg[(2\mu^2-5m^2)\mu\sqrt{\mu^2-m^2}\non[2ex]
&&+3m^4\ln\frac{\sqrt{\mu^2-m^2}+\mu}{m}\Bigg]\,.
\eea
Here we have divided by the factor $N_f{\cal N}$ to use the same dimensionless 
form as dictated by the holographic setup. 
The corresponding (dimensionful) lepton density is 
\be
n_\ell = n(m_e,\mu_e)+n(m_\mu,\mu_\mu) \, ,
\ee
with 
\be
n(m,\mu) \equiv \Theta(\mu-m)\frac{(\mu^2-m^2)^{3/2}}{3\pi^2} \, .
\ee
We shall require equilibrium with respect to the electroweak processes of beta decay and electron capture, $n\to p+e+\bar{\nu}_e$, $e+p\to n+ \nu_e$ and the purely leptonic processes $e\to \mu+\bar{\nu}_\mu+\nu_e$, $\mu\to e+\bar{\nu}_e+\nu_\mu$. At zero temperature this implies the following conditions  for the chemical potentials ("beta equilibrium"), 
\be
\mu_\mu = \mu_e \, , \qquad \mu_e+\mu_p = \mu_n+\mu_\nu \, ,
\ee
where the neutron and proton chemical potentials are given in terms of baryon and isospin chemical potentials by
\be
\mu_n = \mu_B+\mu_I \, , \qquad
\mu_p = \mu_B-\mu_I \, . \label{mupmun}
\ee
We shall assume that neutrinos leave the system once they are created, i.e., their mean free path is of the order of or larger than the size of the star, which is a good approximation except for the very early stages in the life of the star or for mergers. Therefore we set $\mu_\nu=0$, and 
beta equilibrium yields
\be \label{mueI}
\mu_e = 2\mu_I \, .
\ee
In writing down the neutron and proton chemical potentials (\ref{mupmun}) we have interpreted the two isospin components in the space spanned by $\mathbb{1}$ and $\sigma_3$ as neutron and proton contributions. This seems natural, but we should keep in mind that our formalism does not exhibit actual neutron and proton states,  as discussed at the beginning of this section. Within this identification, the  proton and neutron densities are given by 
\be
n_B = n_n+n_p \, , \qquad  n_I = n_n-n_p \, . 
\ee
 Then, assigning the electric charges 0 and $+1$ to the neutron and proton components, local charge neutrality  $n_p=n_e+n_\mu$ can be written as
\be
\frac{\bar{n}_B-\bar{n}_I}{2}-\bar{n}_\ell=0 \, ,
\ee
where the dimensionless lepton density $\bar{n}_\ell$ is related to $n_\ell$ also by the factor given in Eq.\ (\ref{nBbar}). 

So far, the entire setup is taken -- with some notational adjustments for our purposes -- from Ref.\ \cite{Kovensky:2021ddl}. For our application to compact stars 
we also need to compute the energy density,
\bea
\epsilon &=& \Omega +\mu_n n_n+\mu_pn_p+\mu_en_e+\mu_\mu n_\mu \non[2ex]
&=& \Omega+\mu_n(n_n+n_p) \non[2ex]
&=& \epsilon_0 \bar{\epsilon}  \, ,
\eea
where, in the second line, beta equilibrium and charge neutrality have been used, and in the third line we have introduced the dimensionless energy density
\be
\bar{\epsilon} = \bar{\Omega}+(\bar{\mu}_B+\bar{\mu}_I)\bar{n}_B \, ,
\ee
and the corresponding dimensionful factor 
\bea \label{eps0hom}
\epsilon_0&=& N_f{\cal N} \simeq 1.319\times 10^{4} \lambda_0^3\left(\frac{M_{\rm KK}}{\rm GeV}\right)^4\,\frac{\rm MeV}{{\rm fm}^3} \, .
\eea
With the help of the equations from the previous subsection, we can compute the energy density and the resulting equation of state: for instance, for a given $\bar{\mu}_I$ we solve Eqs.\ (\ref{EOMs}), compute the corresponding $\bar{\mu}_B$, $\bar{n}_B$, $\bar{\Omega}_B$ via Eqs.\ (\ref{muBnI}), (\ref{solom}), and the leptonic part from Eqs.\ (\ref{pell}) and (\ref{mueI}). 

\subsection{Constructing a holographic crust}
\label{sec:crust}

The construction described so far accounts for homogeneous nuclear matter in the interior of a neutron star. The outer layers are, however, expected to have a crystalline structure, which requires a different equation of state. Many previous holographic studies employed crust equations of state (plus a low-density layer of the core) from "traditional" methods and added the holographic part for high densities. This is a sensible way to proceed because the underlying low-density microscopic physics are well understood. Therefore,
at least for the outer part of the crust, one might argue that holography is not needed. Nevertheless, here we construct the entire star from holography, which has the advantage of using a single microscopic approach for the entire star. As a consequence, the crust-core transition -- where uncertainties in the more traditional approaches already become sizable -- can be determined dynamically. 

To this end, we construct the crust as follows. Firstly, we observe that our beta-equilibrated, neutral matter shows a first-order transition from the vacuum to nuclear matter. This allows us to construct a mixed phase in the vicinity of this phase transition, where nuclear matter (plus leptons) is spatially separated from a lepton gas. The construction we use here is often employed at the transition between nuclear matter and quark matter in the core, see for instance Refs.\ \cite{Glendenning:1992vb,glendenning,PhysRevLett.70.1355,Schmitt:2020tac}. The resulting structure of the crust is somewhat simplistic. Most notably, our construction is, at best, an approximation for the outer crust -- clusters of nucleons immersed in an electron gas -- while we shall not attempt to construct an inner crust, where the nucleon clusters coexist with a pure neutron phase, see Ref.\ \cite{Chamel:2008ca} for a review of the physics of the neutron star crust. Another simplification we will make is the use of a single geometric structure, namely spherical bubbles of nuclear matter immersed in the lepton gas. Again, this is realistic for the outer crust, but in the inner crust more complicated structures are expected, in particular in the vicinity of the crust-core transition ("nuclear pasta" \cite{2013NatPh...9..396N,Caplan:2016uvu,Schmitt:2017efp,Pearson:2020bxz}). It is possible within the given holographic model to make further refinements along these directions in future studies. However, given the crude approximation our nuclear matter represents to begin with, it seems questionable to refine the details of the inner crust and include different geometric structures  before more fundamental improvements have been made.

Let us now turn to the calculation of the equation of state of our crustal mixed phase. 
We denote the volume fraction occupied by the leptonic phase with $\chi \in [0,1]$, such that the volume fraction of the baryonic phase (nucleons plus leptons)  is $1-\chi$. 
Then, the conditions for the mixed phase read
\begin{subequations}\label{mixeq}
\bea
0&=& \bar{\Omega}_B \, , \label{Pidentical}\\[2ex]
0&=& (1-\chi)\left(\frac{\bar{n}_B-\bar{n}_I}{2}-\bar{n}_\ell\right) - \chi \bar{n}_\ell \, . \label{globalcharge}
\eea
\end{subequations}
The first equation is the condition that the pressure -- and thus the free energy density -- of the baryonic phase,  $\bar{\Omega}_B+\bar{\Omega}_\ell$, be identical to that of the leptonic phase, $\bar{\Omega}_\ell$, where, by construction, 
both phases must have the same electron chemical potential and thus the same $\bar{\mu}_I$. Consequently, the pressure in both constituents of the mixed phase is identical to the leptonic pressure, and the pressure of the baryonic component must be zero for any baryon density and any proton fraction that occurs in the crust. The second equation (\ref{globalcharge}) is the global neutrality condition (allowing each phase
on its own to be electrically charged). 

In order to solve these coupled equations for $\chi$ and $\bar{n}_B$, we first fix  a value for $\bar{\mu}_I$. Then, with the help of (\ref{muBnI2}) and (\ref{solom}) we write  $\bar{n}_I$ and $\bar{\Omega}_B$ as results of a numerical routine that involves solving the differential equations (\ref{EOMs}), which in turn also depend on $\bar{n}_B$. 
Inserting this routine into Eqs.\ (\ref{mixeq}) then allows us to solve them for 
$\chi$ and $\bar{n}_B$. Afterwards, we can compute $\bar{\mu}_B$ from Eq.\ (\ref{muBnI1}), the pressure $\bar{P}_{\rm mix}=-\bar{\Omega}_\ell$, and the energy density 
\be
\bar{\epsilon}_{\rm mix} = (1-\chi)[\bar{\Omega}_\ell+(\bar{\mu}_B+\bar{\mu}_I)\bar{n}_B]+\chi \bar{\Omega}_\ell \, .
\ee
Repeating this procedure for many values of $\bar{\mu}_I$, we find the equation of state.

So far our equations know nothing about the particular geometric structure of the mixed phase. This structure becomes important if surface and Coulomb effects are taken into account. We shall do so in the Wigner-Seitz approximation, where the shape of the unit cell is chosen according to the geometric structure of the mixed phase, e.g., spherical for bubbles. For a given volume fraction $\chi$, the size of the Wigner-Seitz cell is then determined by the competition  between surface tension (preferably large unit cells) and electrostatic Coulomb energy (preferably small unit cells). In principle, these effects 
can be included fully dynamically if the interface profiles are calculated, which also includes screening effects automatically \cite{Maruyama:2005vb,Schmitt:2020tac}. Here we proceed with a simple approximation that assumes the 
interfaces to be sharp surfaces, with spatially uniform charge density in either phase, as often used for the quark-hadron mixed phase \cite{glendenning}. A brief summary and derivation of this approximation can be found in Ref.\ \cite{Schmitt:2020tac}; here we simply quote the relevant results.

The cost in free energy from surface and Coulomb effects in Heaviside-Lorentz units is 
\be \label{DOm}
\Delta \Omega = \frac{3}{2}(\rho_1-\rho_2)^{2/3}\Sigma^{2/3}(1-\chi)[d^2f_d(1-\chi)]^{1/3} \, , 
\ee
with the charge densities of the baryonic and leptonic phases, respectively,
\be
\rho_1 = \frac{e\lambda_0^2M_{\rm KK}^3}{3\pi^2}\left(\frac{\bar{n}_B-\bar{n}_I}{2}-\bar{n}_\ell\right) \, , \quad 
\rho_2 = - \frac{e\lambda_0^2M_{\rm KK}^3}{3\pi^2}\bar{n}_\ell \, ,
\ee
where $e=\sqrt{4\pi\alpha}\simeq 0.3$ is the elementary charge, and with 
\be \label{fd}
f_d(\chi) \equiv \left\{\begin{array}{cc} \displaystyle{\frac{(\chi-1)^2}{3\chi}} & \mbox{for} \;\; d=1 \\[2ex]
\displaystyle{\frac{\chi-1-\ln\chi}{4}} & \mbox{for} \;\; d=2 \\[2ex]\displaystyle{\frac{2+\chi-3\chi^{1/3}}{5}} & \mbox{for} \;\; d=3 
\end{array}\right. \, ,
\ee
where $d$ is the co-dimension of the geometric structure, $d=1$ for slabs, $d=2$ for rods, and $d=3$ for bubbles. 
We have chosen $\chi$ in Eq.\ (\ref{DOm}) such that it corresponds to the volume fraction of the phase in the {\it outer} region of the Wigner-Seitz cell, i.e., if we are interested in bubbles of the baryonic phase immersed in the lepton gas, $\chi$ is the volume fraction of the lepton gas, as in Eq.\ (\ref{mixeq}).
Moreover, $\Sigma$ in Eq.\ (\ref{DOm}) is the surface tension, which, in the given step-like approximation of the interface profiles, is simply an external parameter. For simplicity we shall assume $\Sigma$ to be independent of baryon and isospin density. As a benchmark, it is useful to keep in mind that the empirical value for symmetric nuclear matter at saturation is about $\Sigma\simeq 1 \, {\rm MeV}/{\rm fm}^2$  \cite{Hua:2000gd,Drews:2013hha}   and that the surface tension in the inner crust is expected to become smaller with increasing neutron excess \cite{Douchin:1998fe}. Due to our simple construction of the crust and since we shall not adopt a density-dependent surface tension, it makes sense for us to 
vary $\Sigma$ around its empirical value and check the dependence of our results on this variation.  

For the practical calculation we write the dimensionless free energy cost as 
\bea \label{DOmdimless}
&&\Delta\bar{\Omega} = \frac{3}{4N_c}\left[\frac{6\pi^2 A^4e^2(\bar{n}_B-\bar{n}_I)^2}{\lambda_0^5}\right]^{1/3}(1-\chi)\non[2ex]
&&\times  [d^2f_d(1-\chi)]^{1/3}\left(\frac{\Sigma}{{\rm MeV}/{\rm fm}^2}\right)^{2/3}\left(\frac{M_{\rm KK}}{{\rm MeV}}\right)^{-2} \, , \hspace{0.5cm}
\eea
with the numerical constant $A\simeq 197.327$. This expression can be used to insert the holographic results from Eqs.\ (\ref{mixeq}). Then, the equation of state including surface and Coulomb effects is obtained from the pressure $\bar{P}_{\rm mix,sur}=\bar{P}_{\rm mix}-\Delta\bar{\Omega}$ and the energy density $\bar{\epsilon}_{\rm mix,sur} = \bar{\epsilon}_{\rm mix}+\Delta\bar{\Omega}$.  

\subsection{Thermodynamics and equation of state}
\label{sec:thermo}

We now evaluate the equations of the previous subsections numerically and discuss the thermodynamic properties of our system. The coupling with gravity and the resulting properties of the holographic stars will be discussed separately in Sec.\ \ref{sec:NSs}.

For an explicit evaluation we need to choose values for $\lambda$ and $M_{\rm KK}$. In the present approach -- and in contrast to the pointlike approximation --  the equations of motion (\ref{EOMs}) already depend on $\lambda$ explicitly; there is no rescaling of the variables that eliminates $\lambda$. The equations of motion do not depend explicitly on $M_{\rm KK}$, but this scale enters when we add the lepton gas (due to the lepton masses) and when we compute the surface effects in the mixed phase (due to the surface tension). 
In the present section we shall work with $\lambda=10$ and $M_{\rm KK}=949\, {\rm MeV}$. This particular choice is not very crucial for now because 
the main purpose of this section is to 
point out qualitative properties and establish our construction of the crust. In Sec.\  \ref{sec:NSs} we shall discuss several parameter sets and also present a systematic study in the $\lambda$-$M_{\rm KK}$ plane. To put our parameter choice into context, we have listed three parameter fits in Table \ref{tab:para}, two of which are taken from the literature and one that is obtained by computing the saturation properties of symmetric nuclear matter within the present setup. This table shows that already without any astrophysical constraints the current version of the model is 
too simplistic to account for correct QCD vacuum properties and properties of nuclear matter at the same time. We will further discuss this tension in the parameter space  in Sec.\ \ref{sec:NSs}.

\begin{table}[]
    \centering
    \begin{tabular}{|c||c|c|}
    \hline
       $\;\;$ fit to $\;\;$  & $\;\;\lambda\;\;$ & $\;\;M_{\rm KK}\;\;$ \\
    \hline\hline
        $f_\pi$, $m_\rho$  & $16.63$ & $949\, {\rm MeV}$\\
    \hline
     $\;\;$ $\sigma$, $m_\rho$ $\;\;$  & $\;\;$$12.55$$\;\;$ & $949\, {\rm MeV}$\\
    \hline
     $n_0$, $E_B$  & $7.09$ & $\;\;$$1000\, {\rm MeV}$$\;\;$\\
    \hline     
    \end{tabular}
    \caption{Parameter fits in the confined geometry with maximally separated flavor branes, using pion decay constant $f_\pi$ and rho meson mass $m_\rho$ (original work by Sakai and Sugimoto \cite{Sakai:2004cn,Sakai:2005yt}), including the QCD string tension $\sigma$ for a fit with large-$N_c$ lattice data (used in a study of glueball decay rates \cite{Brunner:2015oqa}), and using saturation density $n_0=0.153\, {\rm fm}^{-3}$ and binding energy $E_B=-16\, {\rm MeV}$ within the homogeneous ansatz for baryonic matter employed in this paper. 
    }
    \label{tab:para}
\end{table}

\begin{figure} [t]
\begin{center}
\includegraphics[width=\columnwidth]{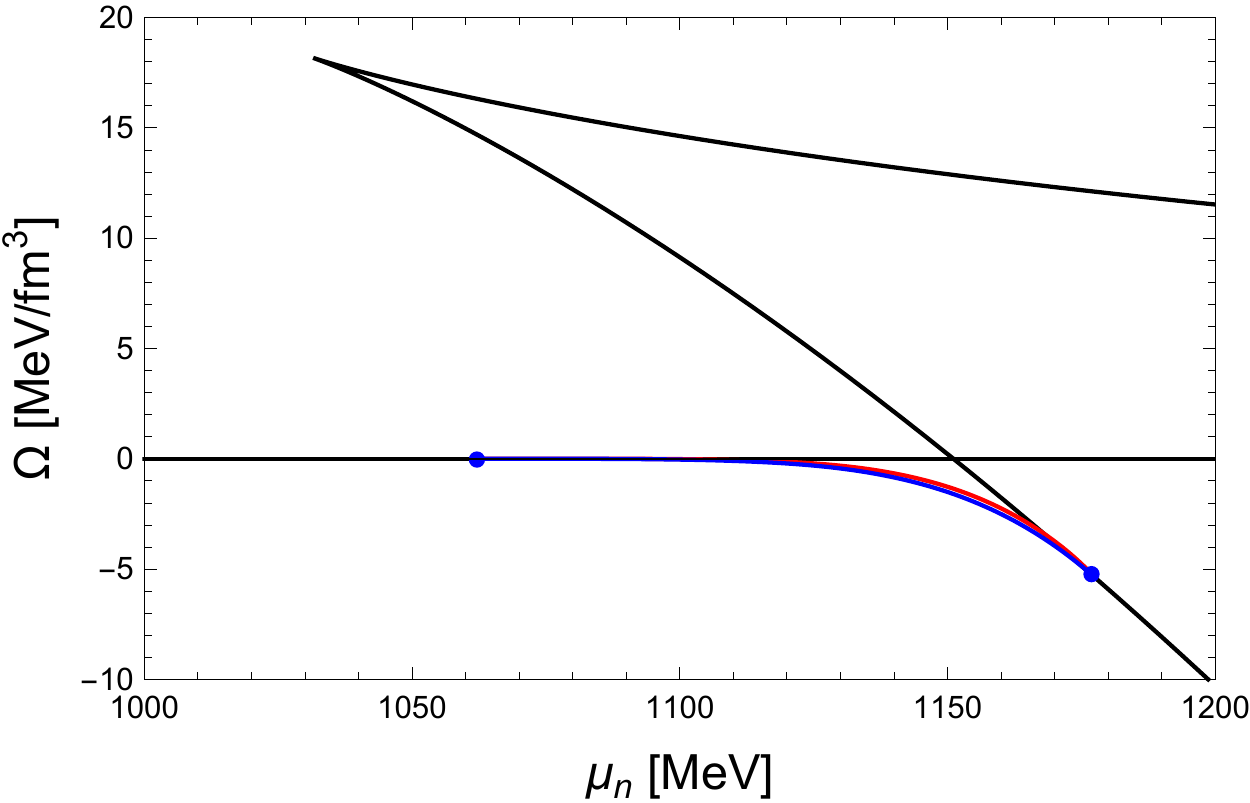}
\includegraphics[width=\columnwidth]{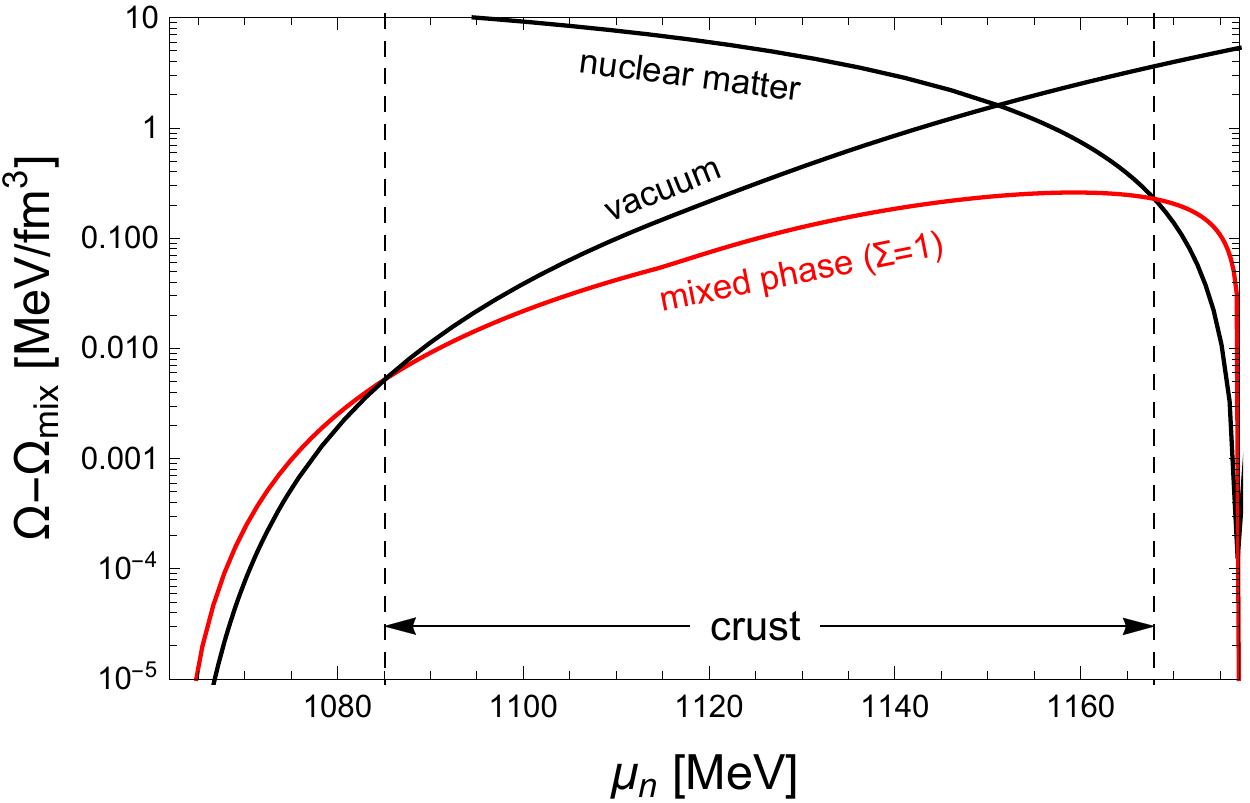}
\caption{{\it Upper panel:} Free energy densities as a function of the neutron chemical potential: pure beta-equilibrated, locally charge neutral baryonic matter (black), globally neutral mixed phase of baryonic matter and a lepton gas without (blue) and with (red) surface and Coulomb effects, where the surface tension is set to $\Sigma=1\,{\rm MeV}/{\rm fm}^2$. The horizontal black line $\Omega=0$ is the free energy density of the vacuum. {\it Lower panel:} Free energy densities relative to that of the mixed phase without surface and Coulomb effects $\Omega_{\rm mix}$. The mixed phase ("crust") is the favored phase between the two dashed lines. The model parameters used for this plot are $\lambda=10$, $M_{\rm KK}=949\, {\rm MeV}$.}
\label{fig:Omega}
\end{center}
\end{figure}

In the upper panel of Fig.\ \ref{fig:Omega} we present various free energy densities 
as a function of the neutron chemical potential (Eqs.\ (\ref{muBmu}) and (\ref{eps0hom})  have been used to obtain the physical units for the given parameter set). The black, two-valued curve is the result for the pure nuclear matter phase, i.e., homogeneous, beta-equilibrated, locally charge neutral matter, including a metastable/unstable branch with positive free energy density. The other black line is the vacuum, $\Omega=0$. If we were to ignore the other curves, there would be a first-order phase transition from the vacuum to nuclear matter.  The blue curve, with endpoints indicated by the dots, is the free energy density for the mixed phase without surface and Coulomb effects. Where it exists it has lower free energy than the pure phases and thus it replaces the first-order transition with two transitions where the baryon density is continuous -- vacuum/mixed phase and mixed phase/nuclear matter. The low-density end of the mixed phase consists of a lepton gas whose volume fraction $\chi$ approaches 1 and whose density approaches 0 and a baryonic phase whose volume fraction $1-\chi$ approaches 0 with $n_B$ remaining nonzero (such that the spatially averaged density  $\langle n_B\rangle=(1-\chi)n_B$ goes to 0). The  high-density end of the mixed phase 
can be found by solving Eqs.\ (\ref{mixeq}) at $\chi=0$ for $\bar{n}_B$ and $\bar{\mu}_I$. 

\begin{figure} [t]
\begin{center}
\includegraphics[width=\columnwidth]{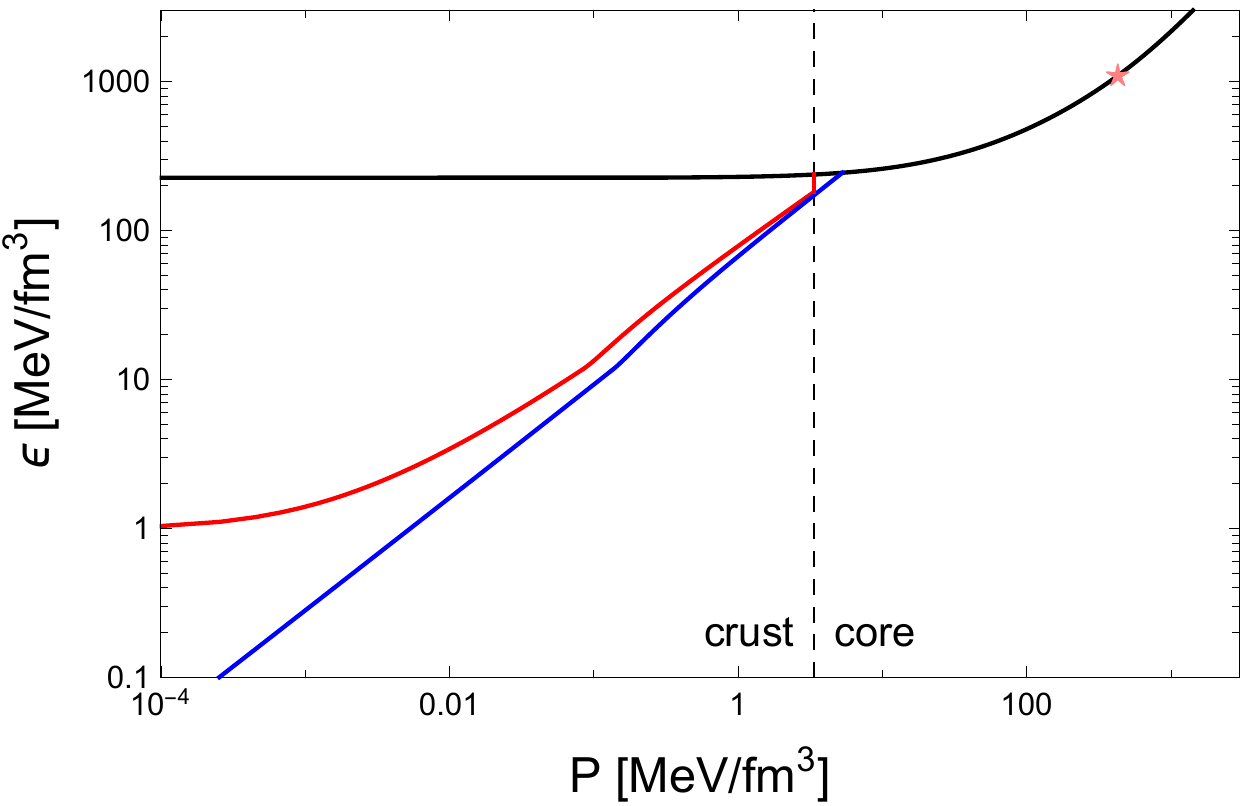}
\includegraphics[width=\columnwidth]{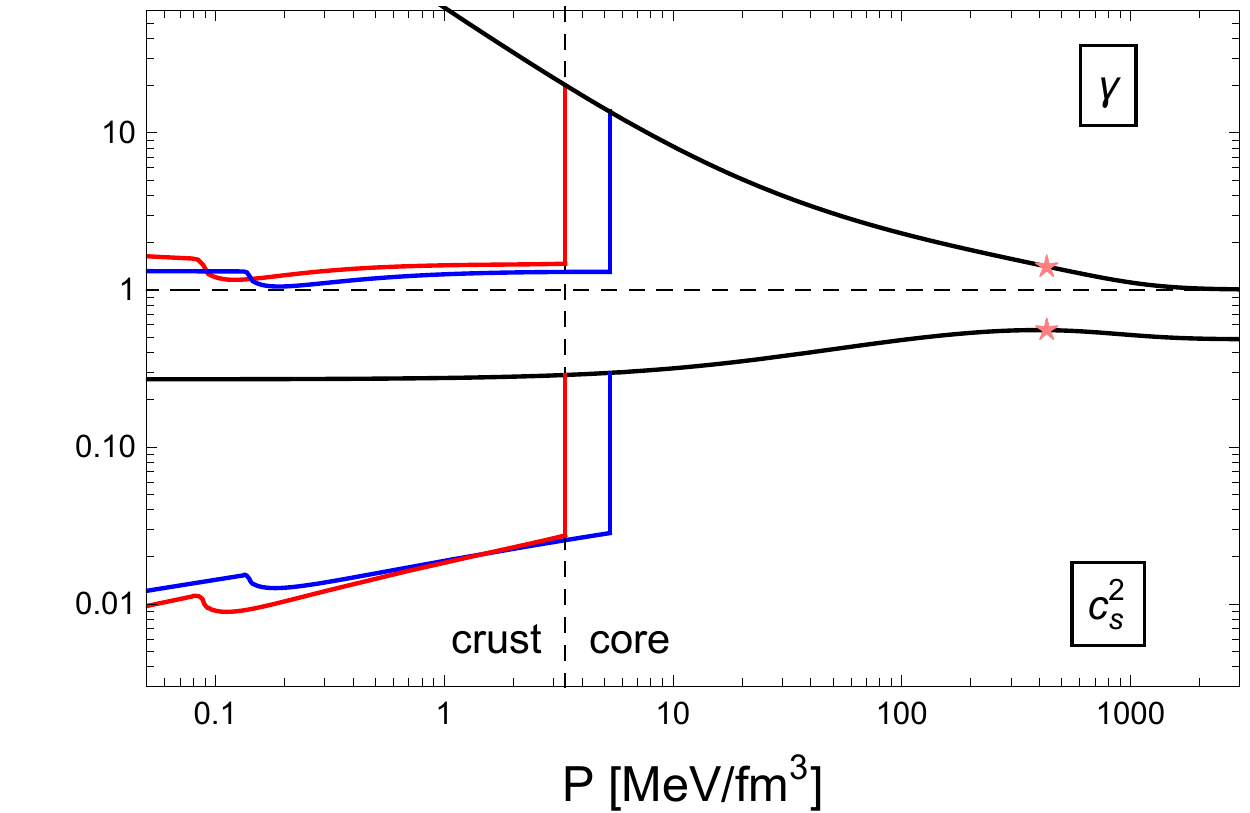}
\caption{Equation of state (upper panel), speed of sound squared, and adiabatic index (lower panel) for the parameters of Fig.\ \ref{fig:Omega}, showing the same phases, i.e., homogeneous nuclear matter (black), mixed phase with (red) and without (blue) surface and Coulomb effects.  The vertical dashed line indicates the crust-core transition, and the star indicates the center of the most massive possible star, computed in Sec.\ \ref{sec:NSs}. The cusp in the curves for the mixed phases is due to the onset of muons. }
\label{fig:epsP}
\end{center}
\end{figure}

The blue curve represents the unphysical form of the mixed phase because surface and Coulomb effects are not included. The result of these effects is shown by the red curve, which is computed with the help of the energy cost (\ref{DOmdimless}), where we have set $d=3$ -- considering spherical nuclear matter bubbles immersed in a lepton gas
-- and $\Sigma=1\, {\rm MeV}/{\rm fm}^2$. Since the 
red curve is barely distinguishable from the blue curve in the upper panel, we plot the difference of all free energy densities relative to the blue curve in the lower panel. As expected, the added energy cost decreases the interval in $\mu_n$ where the mixed phase is preferred, but it still survives. We also see that the transition between the vacuum and the mixed phase as well as between the mixed phase and homogeneous baryonic matter is now of first order again (the slope of the free energy curves changes at the transition points). Anticipating our application of these results to compact stars, the low-density transition point corresponds to the surface of the star, while the high-density transition point corresponds to the crust-core transition inside the star.
We have indicated these transitions by vertical dashed lines. 

\begin{figure} [t]
\begin{center}
\includegraphics[width=\columnwidth]{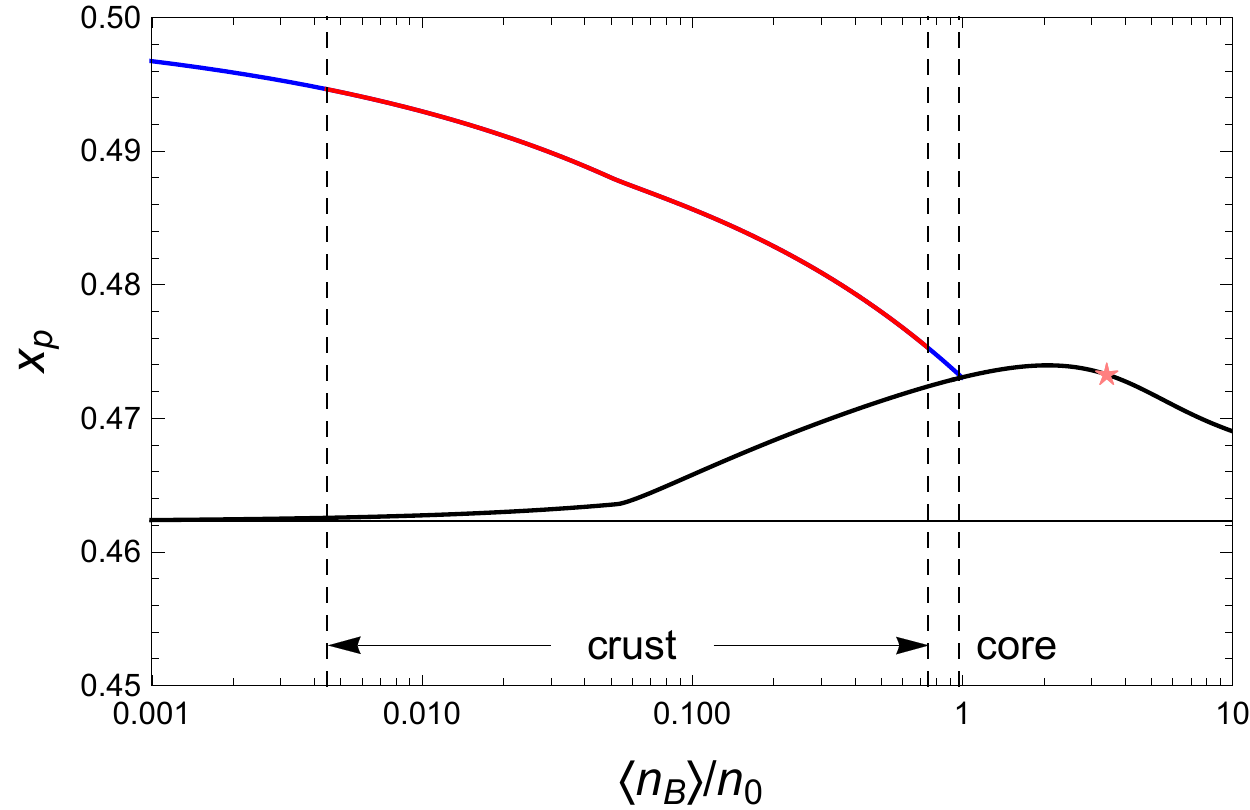}
\caption{Proton fraction as a function of baryon density normalized to the onset density of symmetric nuclear matter, for the parameters of Figs.\ \ref{fig:Omega} and \ref{fig:epsP}. In the mixed phases (red and blue), the baryon number used here is the spatially averaged number $\langle n_B\rangle = (1-\chi) n_B$ (not the baryon number within the clusters). The result of the homogeneous baryonic phase (black solid) approaches the analytic low-density result from Ref.\ \cite{Kovensky:2021ddl} (thin horizontal line). At the surface of the star, $\langle n_B\rangle\simeq 4.4\times 10^{-3}\,n_0$, while at the crust-core transition the density jumps from $\langle n_B\rangle\simeq 0.74 \,n_0$ in the mixed phase to
$n_B\simeq 0.97\, n_0$ in the uniform phase (close to but not exactly at the intercept of black and blue curves).   As in Fig.\ \ref{fig:epsP}, the star corresponds to the center of the most massive star, where 
we find $n_B\simeq 3.45 \,n_0$.}
\label{fig:xP}
\end{center}
\end{figure}

The corresponding equation of state is shown in the upper panel of Fig.\ \ref{fig:epsP}. The fact that the realistic mixed phase (red) has nonzero energy density at zero pressure confirms that there is a (very weak) first-order transition between the vacuum and our crust. Also, the (small) jump in $\epsilon$ at the crust-core transition clearly shows that this transition is of first order. The speed of sound squared and the adiabatic index
\be
\gamma = \frac{\epsilon}{P}\frac{\partial P}{\partial \epsilon} 
\ee
are shown in the lower panel. The adiabatic index has been used 
as a criterion for the distinction between nuclear and quark matter \cite{Annala:2019puf}. Since weakly interacting quark matter has a smaller $\gamma$ than low-density nuclear matter, the value $\gamma=1.75$ was, somewhat arbitrarily, chosen as a value for the transition in Ref.\ \cite{Annala:2019puf}, where a set of parameterized equations of states are used and thus there is no microscopic criterion for this potentially smooth transition. We find that our holographic nuclear matter persists down to the conformal value $\gamma=1$ and even in the center of the star we find values as small as $\gamma\simeq 1.4$. 

In the low-density regime, we see that both $c_s^2$ and $\gamma$ show a curious structure  within the mixed phase (with a close  look a corresponding cusp-like structure can be seen in the equation of state). This is due to the onset of muons, i.e., at that point the electron chemical potential becomes larger than the muon mass. Translated to the stars we shall discuss in the next subsection, this means that muons do appear in the inner part of the crust within our approximation, and not only in the core. The reason of this surprisingly "early" appearance of muons is the large proton fraction $x_p=n_p/n_B$, which can be traced back to the large symmetry energy of our holographic matter, as discussed at the beginning of Sec.\ \ref{sec:beta}. Since, within our holographic approach, it is energetically very costly to move far away from symmetric nuclear matter, the system decides to keep the proton fraction high, which results in a large electron chemical potential and thus in an early onset of muons. In Sec.\ \ref{sec:NSs} we shall, for comparison, also consider the case where muons are omitted altogether and we find that our main conclusions are not very sensitive to their presence. 

It is thus instructive to compute the proton fraction directly. We show the result as a function of baryon density in Fig.\ \ref{fig:xP}. Here we have normalized the baryon density by the saturation density $n_0$ of symmetric nuclear matter. Within the given parameters, we find $\bar{n}_0\simeq 0.089$, which corresponds to $n_0\simeq 0.21\, {\rm fm}^{-3}$ (as discussed above, the present parameters are not chosen to reproduce the physical saturation density exactly). While the overall magnitude of the proton fraction is larger than expected in 
real-world neutron stars, it is worth noticing the qualitative behavior: as expected from a realistic crust, we start off with nearly symmetric matter at very low (spatially averaged) densities and produce more neutron-rich matter as we move towards the core. Within the core, the proton fraction increases before it starts to decrease again at ultra-high densities. This decrease is relevant for the most massive stars, as we have indicated in the figure.

\subsection{Holographic neutron stars}
\label{sec:NSs}

\subsubsection{Mass and radius}

\begin{figure} [t]
\begin{center}
\includegraphics[width=\columnwidth]{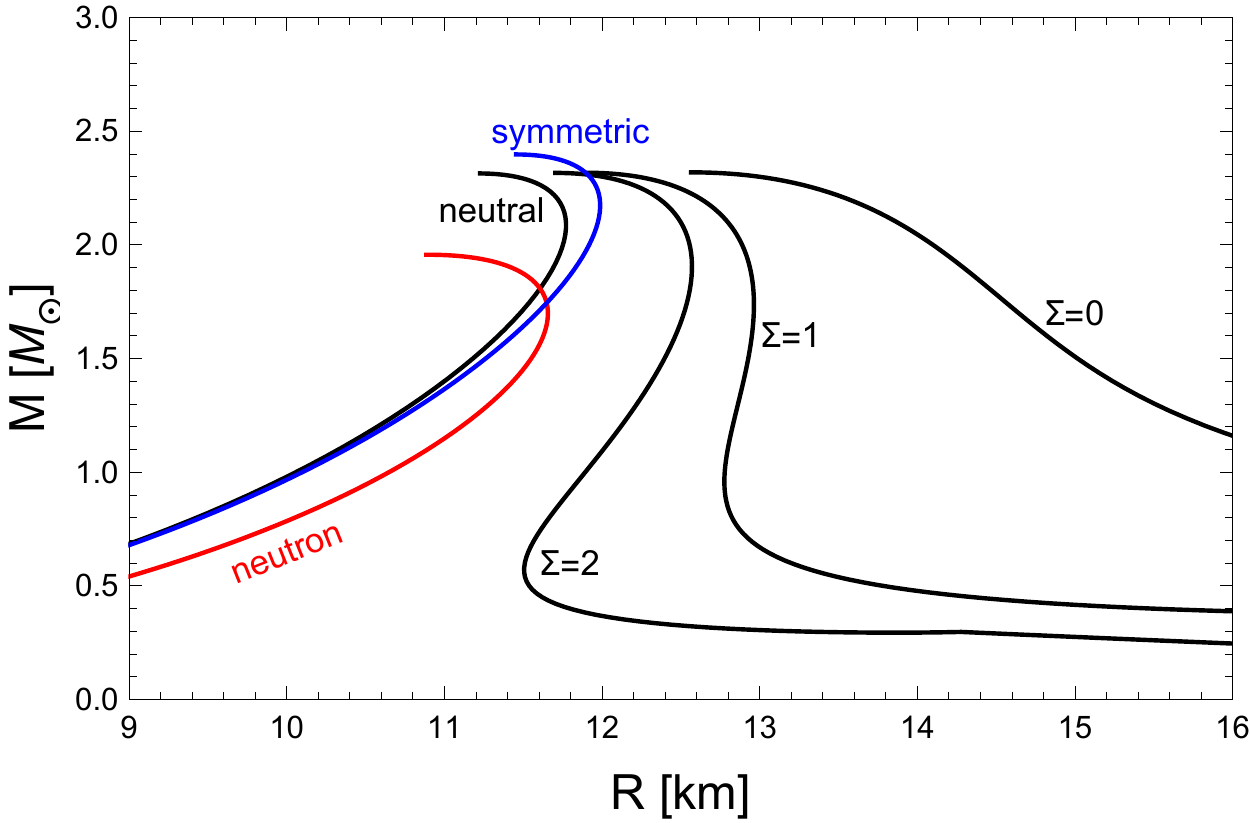}
\caption{Mass-radius curves for pure neutron matter (red), symmetric nuclear matter (blue) (both without crust) and beta-equilibrated, neutral matter (black) without crust and with crust where the surface tension is $\Sigma=0,1,2\, {\rm MeV}/{\rm fm}^2$. For all curves, $\lambda=10$, $M_{\rm KK}=949\, {\rm MeV}$.}
\label{fig:6MRcurves}
\end{center}
\end{figure}

We now insert the equation of state and the speed of sound into the TOV equations and the equation needed for the tidal deformability (\ref{TOVs}). The natural choice of the dimensionless pressure and energy density from Eq.\ (\ref{tilde}) is 
$\hat{P} = \bar{P}$, $\hat{\epsilon}=\bar{\epsilon}$, such that the scale for the energy density $\epsilon_0$ in Eqs.\ (\ref{tilde}) and (\ref{AB1}) is given by Eq.\ (\ref{eps0hom}). (This scale differs from the one used in the pointlike approximation by the factor $\ell^7$; in the present approach $\ell$ does not appear explicitly because the asymptotic separation of the flavor branes is fixed to be antipodal from the beginning.) From this scale and Eq.\ (\ref{AB1}) we determine the scales for mass and radius
\begin{subequations}
\bea
M_0 &\simeq& 1.445\, \lambda_0^{-3/2}\left(\frac{M_{\rm KK}}{{\rm GeV}}\right)^{-2} \, M_{\odot} \, , \\[1ex]
r_0 &\simeq& 2.135\, \lambda_0^{-3/2}\left(\frac{M_{\rm KK}}{{\rm GeV}}\right)^{-2} \, {\rm km} \, . 
\eea
\end{subequations}
For now we continue with the parameters of the previous subsection, $\lambda=10$, $M_{\rm KK}=949\, {\rm MeV}$. Several mass-radius curves with these parameters are shown in Fig.\ \ref{fig:6MRcurves}. In this and the following plots, segments of the curves representing stars unstable with respect to radial oscillations have been omitted,  each curve  terminates at its maximal mass.

Let us first focus on the black curves, which are all for beta-equilibrated, charge neutral matter, but which are obtained with different descriptions for the crust. All curves give almost the same maximal mass, and this maximal mass is consistent with astrophysical constraints if we assume that the 2.5 solar mass object in the merger GW190814 is not a neutron star (see Sec.\ \ref{sec:astroconstr} for the list of astrophysical constraints we discuss here and in the following). If the crust is completely ignored, the mass-radius curve bends back to the origin, resulting in a relatively small radius for, say, a 1.4 solar mass star. This behavior is easy to explain since without the crust there is a first-order transition between our holographic nuclear matter and the vacuum. In other words,
 nuclear matter at a certain nonzero density can coexist with the vacuum, and that's what happens at the surface of the thus constructed star. This allows for arbitrarily small and light chunks of matter, where gravity does not play any role, close to the origin of the $M$-$R$ plane, and the curve assumes the typical shape of self-bound stars often encountered for quark stars in simple models (if the crust of the quark star is neglected) \cite{Postnikov:2010yn}. 
 
The other extreme, giving rise to much larger radii, is shown by the curve labeled by $\Sigma=0$, meaning we have included the crust but neglected surface and Coulomb effects
 (blue curves in Figs.\ \ref{fig:Omega} -- \ref{fig:xP}). This results in unphysically large crusts and overall radii. As we have seen in the previous subsection, surface and Coulomb effects reduce the regime where the mixed phase is favored (red curves in Figs.\ \ref{fig:Omega} -- \ref{fig:xP}), and as a consequence give rise to smaller, more realistic crusts. 
 Examples are the two mass-radius curves in between the two extremes. While the surface tension $\Sigma=1\, {\rm MeV}/{\rm fm}^2$ is close to the empirical value for symmetric nuclear matter, we have also included a curve for $\Sigma=2\, {\rm MeV}/{\rm fm}^2$ to illustrate the effect of a variation in the surface tension, which for simplicity we keep constant throughout the crust for each star of a given curve. We already see that our holographic construction not only reproduces reasonable maximal masses but also, in the most realistic version of the crust, is able to reproduce realistic radii. We will come back to this observation below in a more systematic survey of our parameter space. Strictly speaking, the curves with realistic crust also bend back to the origin for extremely small central pressures. The reason is that even in the presence of the crust (if surface and Coulomb effects are taken into account) there is a (weak) first-order transition to the vacuum, see Figs.\ \ref{fig:Omega} and \ref{fig:epsP}. However, this segment of the mass-radius curves is irrelevant for our purposes and thus  not included in the figure.   

\begin{figure} [t]
\begin{center}
\includegraphics[width=\columnwidth]{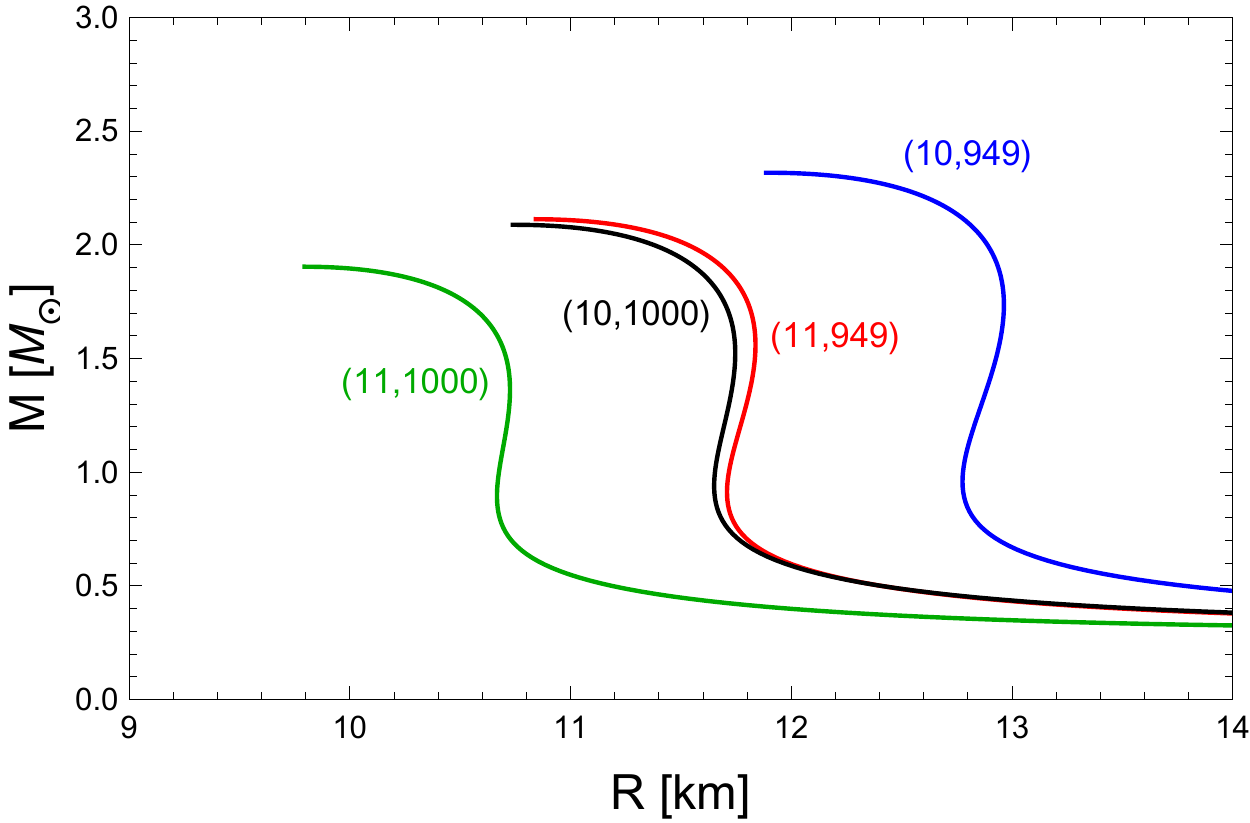}
\caption{Mass-radius curves for four different parameter sets $(\lambda,M_{\rm KK})$, all for beta-equilibrated, neutral matter including a crust with surface tension $\Sigma=1\,{\rm MeV}/{\rm fm}^2$.}
\label{fig:4sets}
\end{center}
\end{figure}

As discussed above, our holographic nuclear matter is plagued by the large-$N_c$ artifact of a very large proton fraction. To get an idea of the size of this effect, we have added two more mass-radius curves in Fig.\ \ref{fig:6MRcurves} where we fix the proton fraction by hand. 
We show the result for pure neutron matter (red) -- defined in our holographic calculation by $n_B=n_I$ -- and for isospin-symmetric nuclear matter $n_I=0$ (blue). Both cases are 
artificial in the sense that the resulting stars are not electrically neutral and not in beta equilibrium. 

We see that, as expected, our isospin-asymmetric stars with 
neutrality and beta equilibrium are better approximated by isospin-symmetric matter than by pure neutron matter. 
The plot suggests that improvements of the holographic model to account for quantized isospin states will  move our mass-radius curve more towards the pure neutron matter curve. This would imply that our current  approximation somewhat overestimates the maximal mass. It is also worth noticing that the mass-radius curve for pure neutron matter shows the same qualitative behavior 
of reaching back to the origin as the other cases without crust. The reason is that 
within our holographic approach even pure neutron matter has a nonzero saturation density at which the pressure is zero. This is in contradiction with results from chiral effective theory, which show that pure neutron matter has positive nonzero pressure for all densities \cite{Hebeler:2013nza}. Given the large-$N_c$ artifacts of our isospin spectrum, this discrepancy is not surprising. 

\begin{figure} [t]
\begin{center}
\includegraphics[width=\columnwidth]{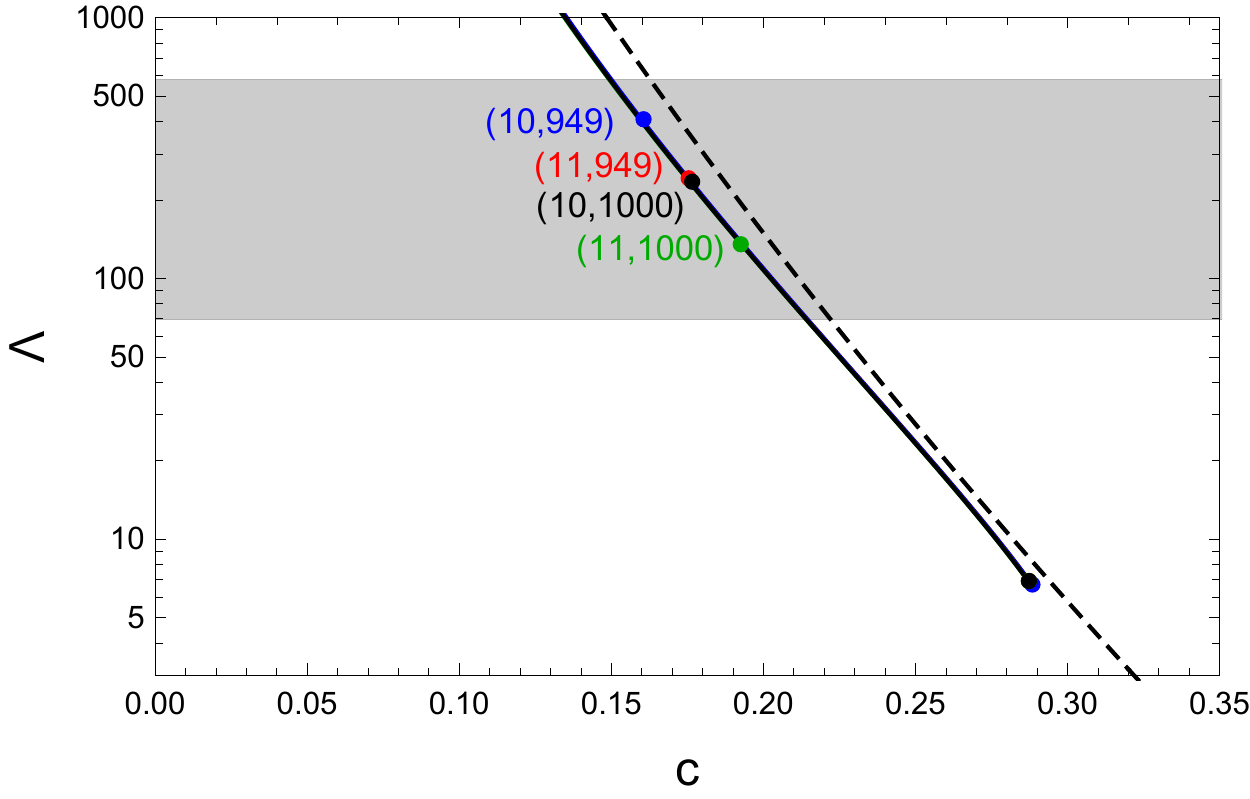}
\includegraphics[width=\columnwidth]{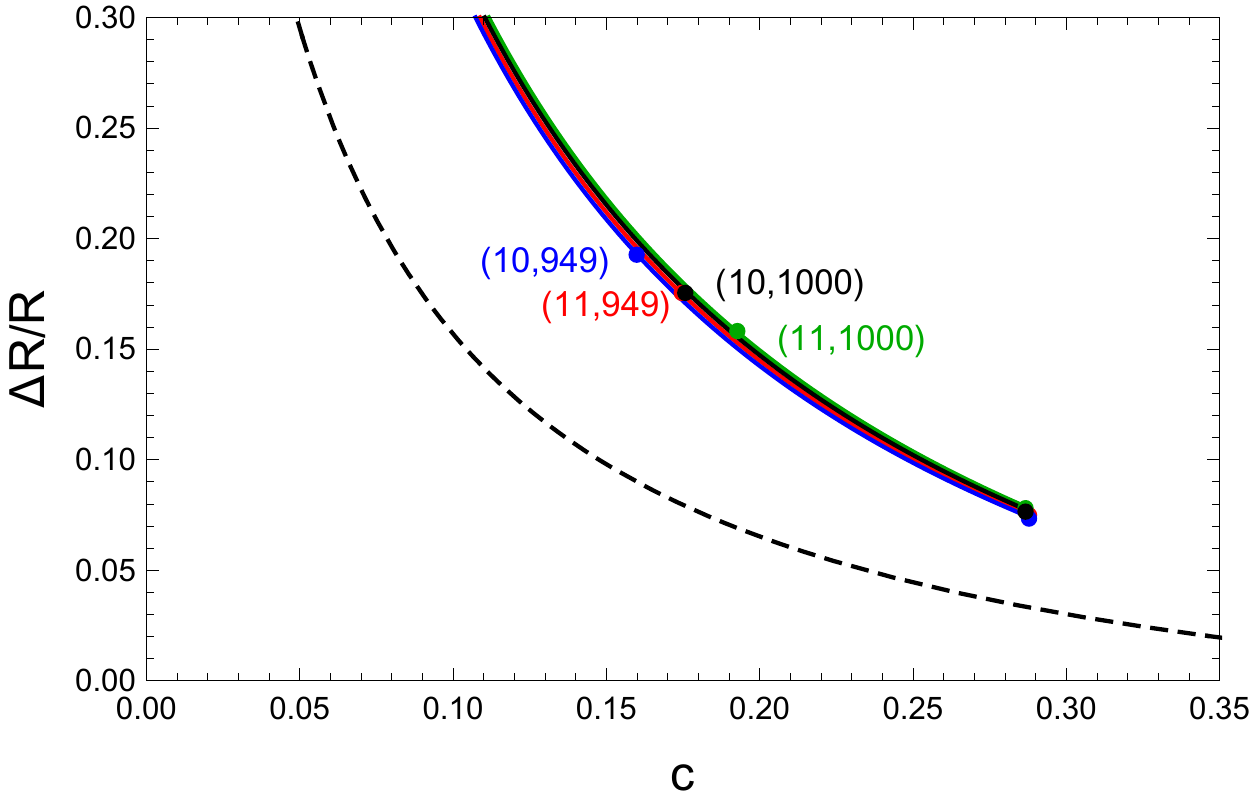}
\caption{{\it Upper panel:} Tidal deformability $\Lambda$ as a function of compactness $c$ for the same four  parameters sets as in Fig.\ \ref{fig:4sets} (all curves are indistinguishable on the given scale). The grey band indicates the constraints from the merger GW170817 for a 1.4 solar mass star and the dots within the band correspond to the tidal deformability of a 1.4 solar mass star $\Lambda_{1.4}$, with colors corresponding to the colors in Fig.\ \ref{fig:4sets}. The dots at the endpoints (hardly distinguishable) mark the most massive stars. The dashed curve is a parameterization constructed to  approximate a large number of equations of state taken from Ref.\ \cite{Yagi:2016bkt}. {\it Lower panel:} Ratio of the crust thickness $\Delta R$ over the radius $R$ of the star as a function of compactness, for the same four parameter sets, again with dots indicating the 1.4 solar mass stars and the most massive stars. Again, the dashed line shows a simple parameterization,  taken from Ref.\ \cite{Samuelsson:2006tt}. }
\label{fig:LamThick}
\end{center}
\end{figure}

In Fig.\ \ref{fig:4sets} we only keep the most realistic version of our model, i.e., including a crust with surface tension $\Sigma=1\, {\rm MeV}/{\rm fm}^2$, but now we study the dependence of our results on the model parameters $\lambda$ and $M_{\rm KK}$. For this discussion we recall the results from Table \ref{tab:para}, which defines a parameter window where we expect the model to approximately reproduce properties of the QCD vacuum and QCD matter. The choice of the previous plots $\lambda=10$, $M_{\rm KK}=949\, {\rm MeV}$ lies within this regime, and in Fig.\ \ref{fig:4sets} we include the mass-radius curves for three additional parameter pairs chosen such that all curves show more or less realistic maximal masses (one of the curves gives a maximal mass slightly below 2 solar masses). The step size of the variation in $\lambda$ and $M_{\rm KK}$ is chosen such that the variation in either variable has a similar effect. For instance, starting from the parameter set used in the previous plots (blue curve), the maximal mass is decreased by roughly the same amount by either increasing $\lambda$ by 1 or increasing $M_{\rm KK}$ by about $50\, {\rm MeV}$.

\subsubsection{Tidal deformability and crust thickness}

We may also check the corresponding tidal deformability $\Lambda$ against the known constraints. This is done in the upper panel of Fig.\ \ref{fig:LamThick}, where we plot $\Lambda$ as a function of the compactness $c$. Here we see that for all four parameter sets 
$\Lambda_{1.4}$ lies in the band given by estimates deduced from the merger GW170817.
Given that they also produce realistic masses, this is in stark contrast to the pointlike approximation of Sec.\ \ref{sec:pointlike}. Remarkably, all curves $\Lambda(c)$ are indistinguishable on the given scale (while the value for $\Lambda_{1.4}$ {\it does} depend visibly on the parameters). This 
apparent universal behavior is not obvious since here, in contrast to the pointlike approximation, the equation of state even in its dimensionless version explicitly depends on $\lambda$ (via the holographic equations of motion) and on $M_{\rm KK}$ (via the lepton masses and the surface tension). We have checked that for 't Hooft couplings of about $\lambda=50$
variations of the curve are visible on the chosen scale, although they are still small.
It is known that different equations of state show an approximately universal behavior 
if certain observables are set in relation. We have, for comparison to our results, included a simple parameterization of the function $\Lambda(c)$ (dashed curve) taken from Eq.\ (78) in Ref.\ \cite{Yagi:2016bkt} that is believed to approximate a wide class of equations of states. (See also Ref.\ \cite{Maselli:2013mva} for a very similar parameterization.)

A similar universal behavior is seen in the lower panel of 
Fig.\ \ref{fig:LamThick}, which shows $\Delta R/R$, where $\Delta R=R-R_{\rm cc}$
is the thickness of the crust, $R_{\rm cc}$ being the radial location of the crust-core transition. Again, the values for $M=1.4\, M_\odot$ are marked by dots. The absolute values for crust thicknesses and radii for the four cases in the order (blue, red, black, green)  are $\Delta R_{1.4} \simeq (2.48,2.06,2.06,1.69)\,{\rm km}$ and $R_{1.4} \simeq (12.9,11.8,11.7,10.7)\,{\rm km}$. These crust thicknesses are somewhat larger than the usually assumed values centered around $1\,{\rm km}$, which, however, are subject to significant uncertainties \cite{Steiner:2014pda,Fortin:2016hny,Grams:2021lzx}. For the most massive stars in each case, also marked by dots, we find the absolute values $\Delta R=(0.88,0.82,0.83,0.77)\, {\rm km}$. Again we show a simple parameterization, taken from Eqs.\ (B6) and (B8) of Ref.\ \cite{Samuelsson:2006tt}, that is believed to be a good approximation independent of the exact behavior of the equation of state. We see that although the qualitative behavior is the same, our results for $\Delta R/R$ are significantly larger than those given by this parameterization.

\begin{figure} [t]
\begin{center}
\includegraphics[width=\columnwidth]{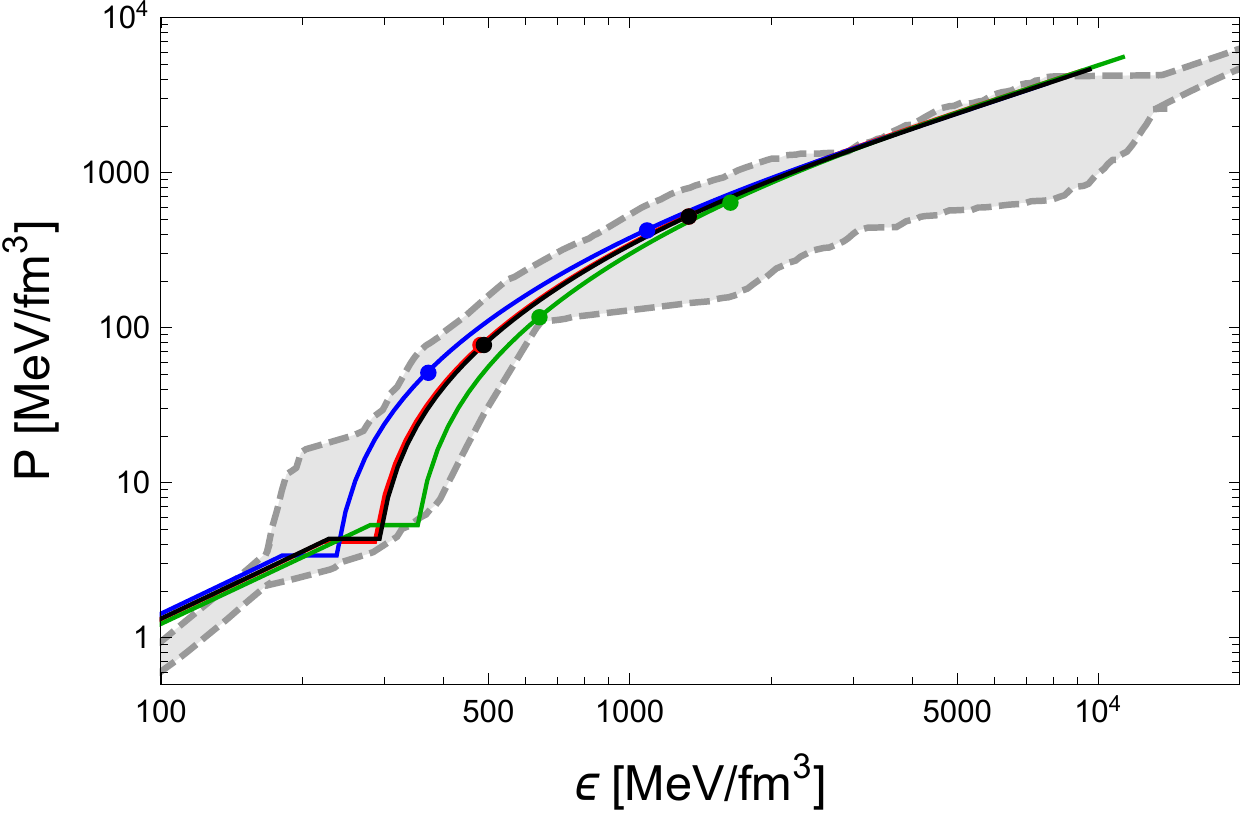}
\caption{Equations of state corresponding to the four parameter sets of Fig.\ \ref{fig:4sets} and \ref{fig:LamThick}, compared with the "allowed" band of Ref.\ \cite{Annala:2017llu}, defined by low-density chiral effective theory, high-density perturbative QCD, and polytropic interpolations between them, constrained by astrophysical observations.
The dots mark the 1.4 solar mass stars and the most massive stars for each equation of state. }
\label{fig:sausage}
\end{center}
\end{figure}

In Fig.\ \ref{fig:sausage}, in order to further put our results into context, we compare the four equations of state used for Figs.\  \ref{fig:4sets} and \ref{fig:LamThick} with the "allowed" band of Ref.\ \cite{Annala:2017llu}\footnote{We thank the authors of Ref.\ \cite{Annala:2017llu} for providing the data of this band.}. At low densities, $n_B \lesssim 1.1\, n_0$, this band contains the results for homogeneous, beta-equilibrated, charge neutral nuclear matter, obtained from a phenomenological extrapolation of pure neutron matter from chiral effective field theory \cite{Tews:2012fj,Hebeler:2013nza}, and shown to behave similarly to a widely used equation of state for the crust \cite{Baym:1971pw,Negele:1971vb}. At extremely high densities, $n_B\gtrsim 40 \,n_0$, the band represents the predictions from perturbative QCD  \cite{Kurkela:2009gj,Fraga:2013qra}. The intermediate regime is constructed from an interpolation using piecewise polytropic equations of state obeying thermodynamic consistency, causality, and the astrophysical constraints $M_{\mathrm{max}}>2 \,M_\odot$ and $\Lambda_{1.4}<580$. We see that our crust deviates from the low-density band, while the intermediate and high-density parts are mostly within the allowed region. The curve  that slightly violates the band at intermediate densities corresponds precisely to the parameter set that does not produce stars above two solar masses, i.e., this behavior is in agreement with the findings of Ref.\ \cite{Annala:2017llu}. All our curves continue within the allowed region even beyond the highest densities in the most massive stars, before they deviate from the band at ultra-high densities (our curves stop where the numerical evaluation becomes problematic). Since our holographic model does not account for asymptotic freedom, it is no surprise that the weak-coupling results are not reproduced. More importantly, we conclude that the asymptotic constraints do not invalidate our results: it is conceivable that our holographic equation of state approximates real-world QCD up to (and even somewhat past) the highest densities in neutron stars, and beyond that, where we do not intend to apply our results anyway, it is possible to connect them to the first-principle QCD calculation without violating any theoretical or astrophysical constraints. 

\begin{figure*} [t]
\begin{center}
\includegraphics[width=\columnwidth]{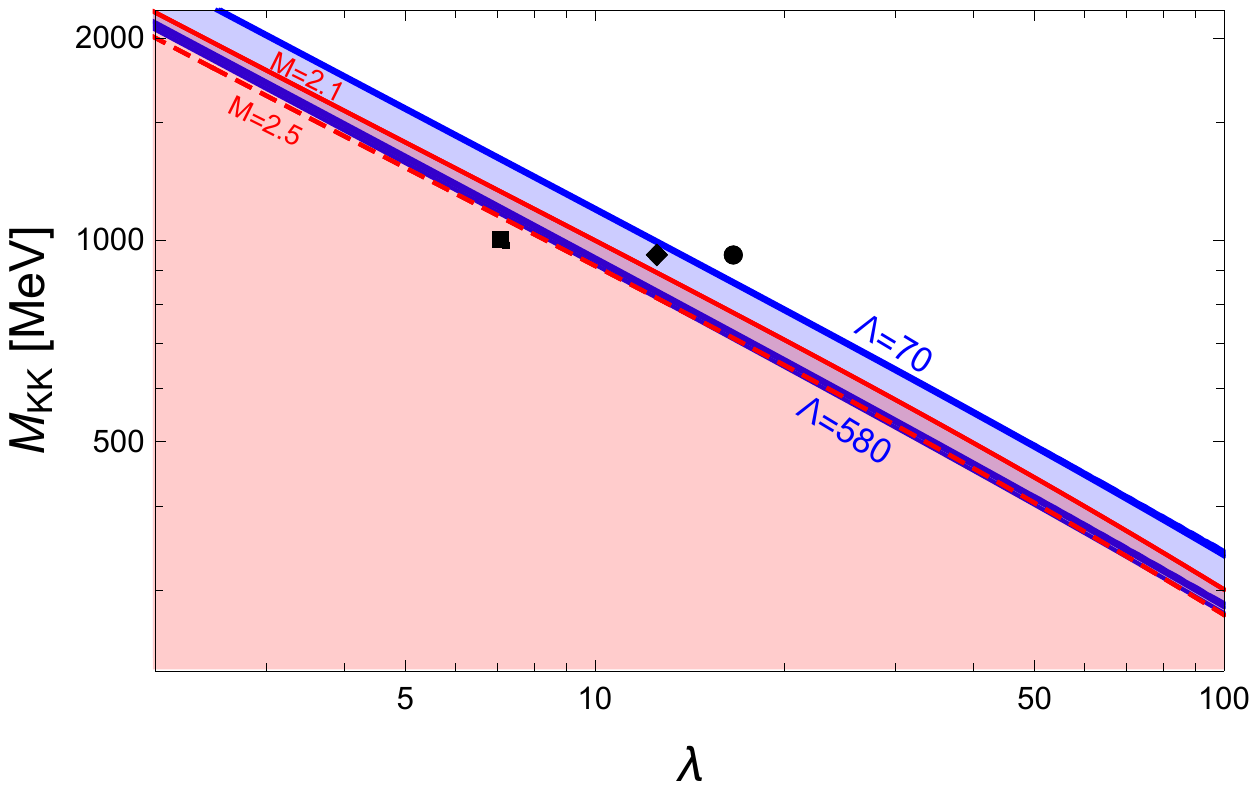}
\includegraphics[width=\columnwidth]{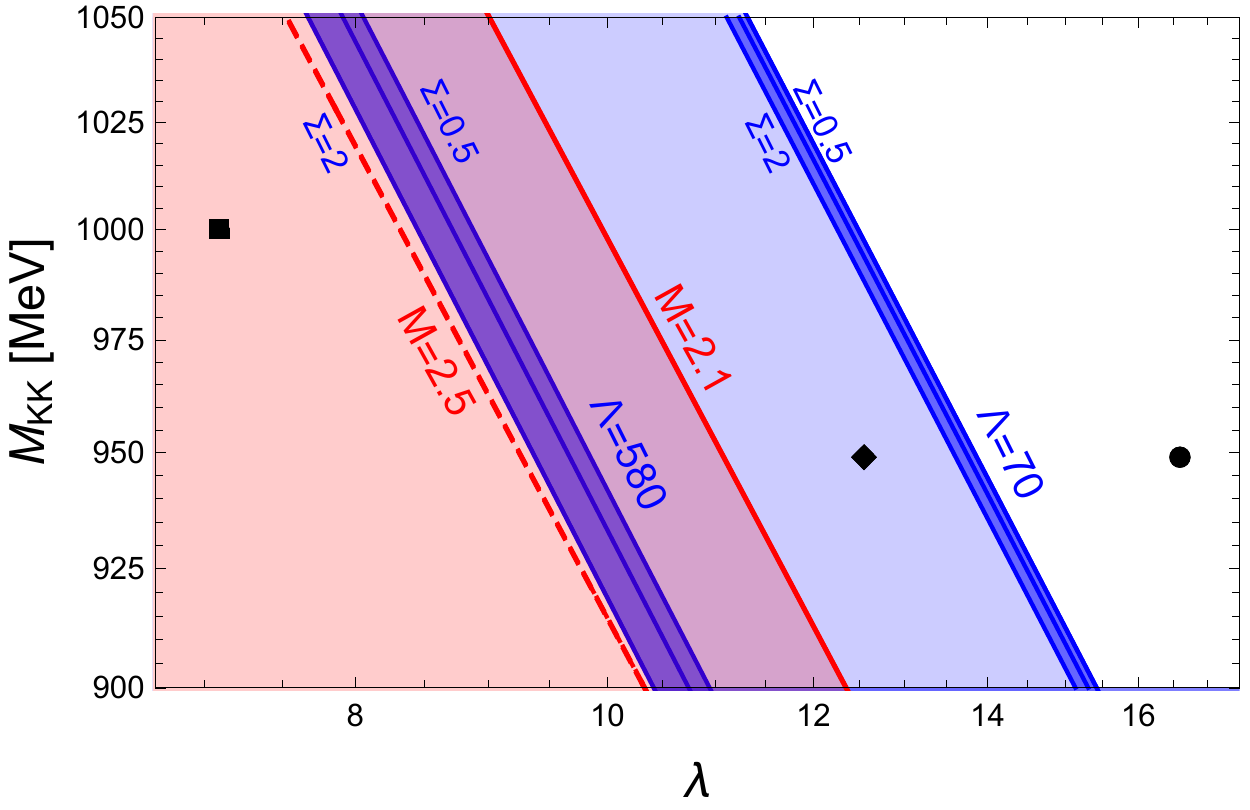}
\caption{Constraints on maximum mass (shaded red for $M_{\rm max}>2.1\,M_\odot$, dashed red line $M_{\rm max}=2.5\,M_\odot$) and deformability of a $1.4\,M_\odot$ star (shaded blue for $70<\Lambda_{1.4}<580)$ in the plane of the model parameters $\lambda$ and $M_{\rm KK}$ on logarithmic scales for surface tensions in the crust  $\Sigma = 0.5,1,2\, {\rm MeV}/{\rm fm}^2$ (differences for different surface tensions only become visible in the zoom-in of the right panel). The dots mark the parameter choices given in Table \ref{tab:para}: fit to $m_\rho$ and $f_\pi$ (circle), to $m_\rho$ and $\sigma$ (diamond), and to the saturation properties of nuclear matter (square). In this plot, muons are neglected for simplicity. }
\label{fig:lamMKK}
\end{center}
\end{figure*}

\subsubsection{Discussion of the parameter space}

Finally, let us discuss our parameter space more systematically. To this end, we determine the regions in the $\lambda$-$M_{\rm KK}$ plane where astrophysical constraints on mass and tidal deformability are obeyed, see Fig.\ \ref{fig:lamMKK}. Before we discuss the results let us briefly explain how this figure was produced. For simplicity, here we have neglected the electron mass (which makes almost no difference for any observables we discuss) and ignored muons (which does make a small difference, but does not affect the main conclusions). With these simplifications, the holographic calculation itself, which is the most time-consuming part of our numerics, becomes independent of $M_{\rm KK}$, and we only have to do it once for each $\lambda$. The energy cost from Coulomb and surface effects
unavoidably induces a dependence on $M_{\rm KK}$. Adding this cost within our approximation to obtain the equation of state is trivial, but we 
have to solve the TOV equations -- which is numerically less demanding than the holographic part of the calculation -- on a suitably fine grid in the $\lambda$-$M_{\rm KK}$ plane for each grid point separately. In this way we can, for each $\lambda$, determine the value $M_{\rm KK}$ needed to obtain a given maximal mass and a given tidal deformability of a 1.4 solar mass star.

As a result, we can identify the windows where the maximal mass is larger than 2.1 solar masses (shaded red) and where  the tidal deformability is within the bounds $70<\Lambda_{1.4}<580$ (shaded blue). For completeness we have also added the contour for the maximal mass of 2.5 solar masses (red dashed curve), having in mind the possibility of an ultra-heavy neutron star in the merger event GW190814 (although a general upper bound for the maximal mass below that value was suggested \cite{Rezzolla:2017aly}). We have also used three different values of the surface tension. This creates a small variation in the deformability and essentially no change in the maximal mass, see right panel for a zoom-in, where the (blue) lines for the three different values of $\Sigma$ become distinguishable. The main observation of this figure is that there is a region in the parameter space where the constraints of the 2.1 solar masses and the tidal deformability can be met, as already seen for selected parameters in the previous plots. However, a 2.5 solar mass star is more difficult to reconcile with the constraint for $\Lambda_{1.4}$, as the red dashed line barely enters the blue area. 

We have indicated the three particular parameter choices of Table \ref{tab:para} in the plot. We see that none of them lies in the overlap region of the red and blue areas. For the interpretation of this result we should keep in mind that they already disagree with each other, i.e., had we looked for a parameter set within the given approximation that can fit "everything" it would not have been necessary to calculate properties of compact stars. 
On the other hand, we may use Table \ref{tab:para} to define a window in parameter space 
in which QCD properties of the vacuum and of nuclear matter at saturation can be reproduced simultaneously at least in an approximate way. It is reassuring that this window overlaps with the one defined by the astrophysical constraints in Fig.\ \ref{fig:lamMKK}. We thus conclude that the Witten-Sakai-Sugimoto model -- even in its simplest version and by fitting only two parameters -- can account approximately for observables in regimes as diverse as the vacuum, nuclear matter at saturation, and basic properties of neutron stars. 
Moreover, the small number of parameters distinguishes our approach 
from many phenomenological models used for dense matter in neutron stars. This allows us also to ask whether we can make some parameter-independent observations, as we shall discuss next.

\begin{figure*} [t]
\begin{center}
\includegraphics[width=\columnwidth]{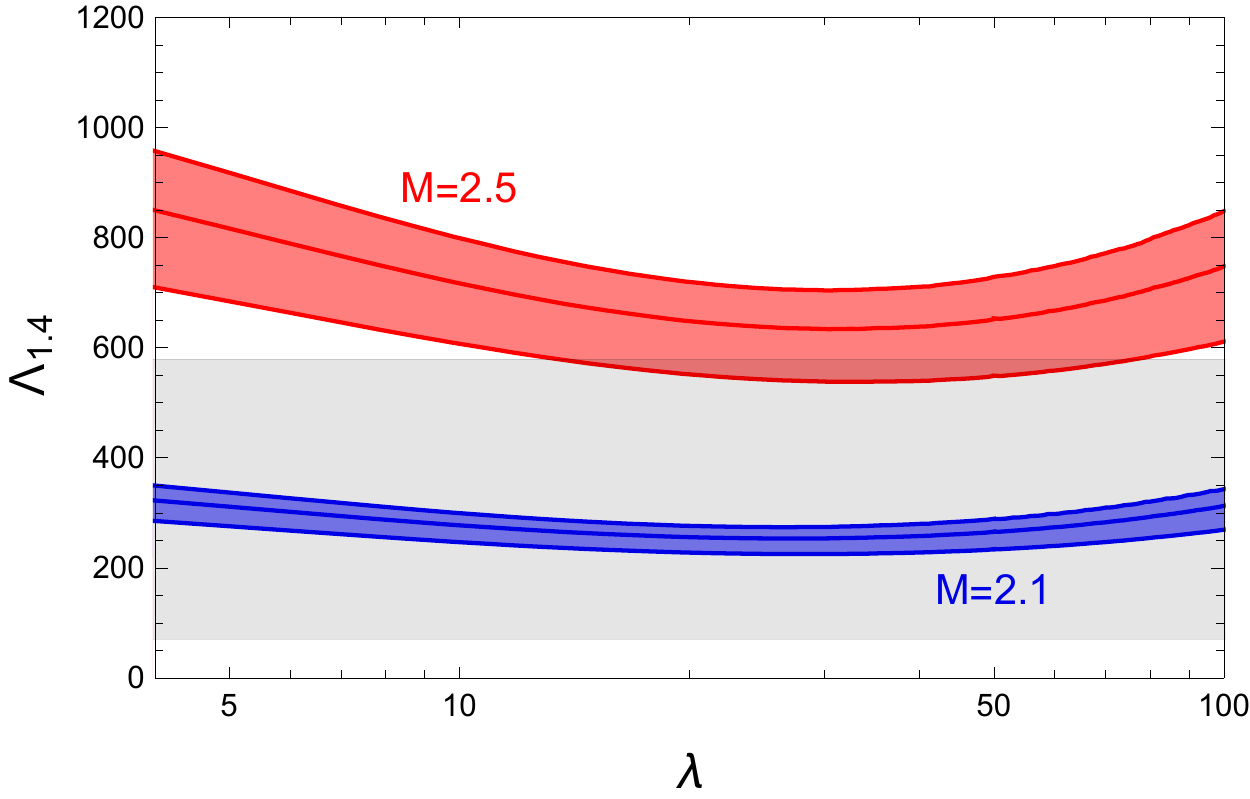}
\includegraphics[width=\columnwidth]{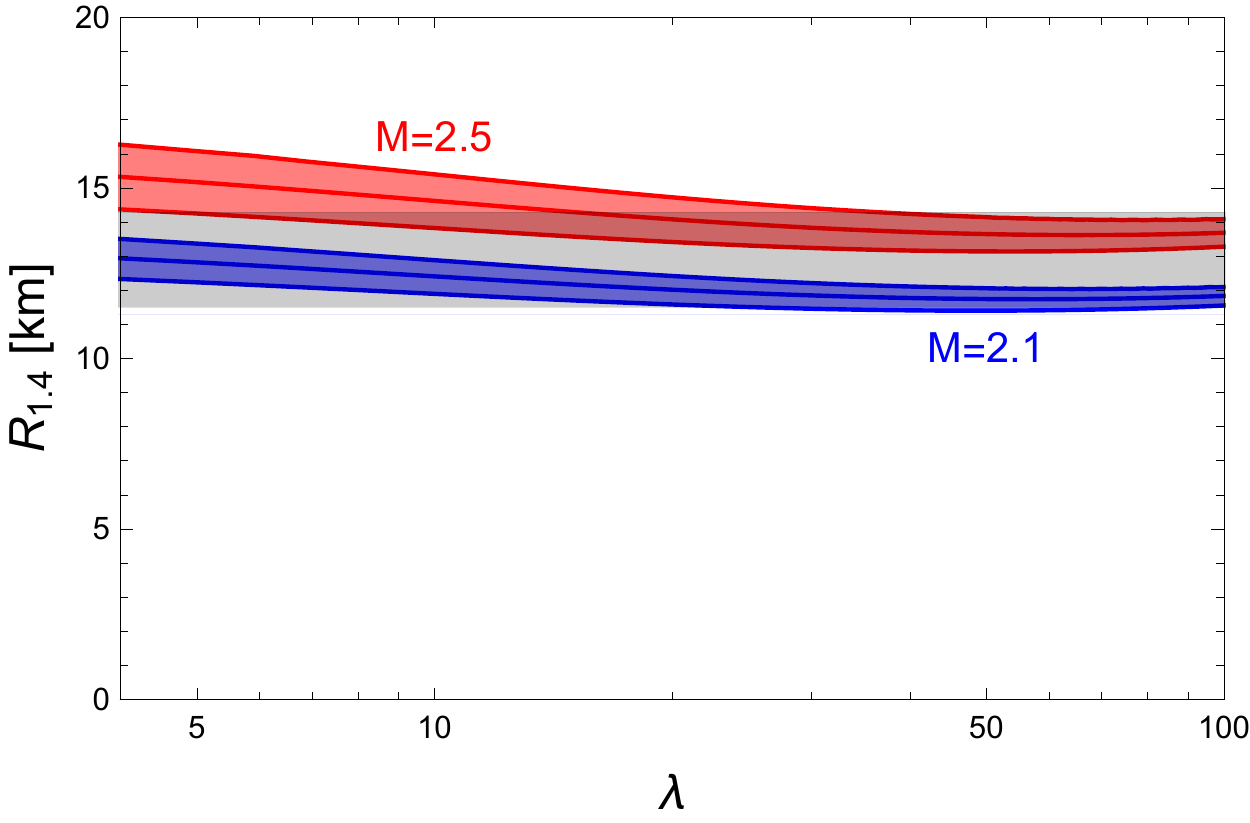}

\caption{Tidal deformability (left panel) and radius (right panel) for a 1.4 solar mass star as a function of the 't Hooft coupling $\lambda$ with $M_{\rm KK}$ adjusted for each $\lambda$ such that the maximal mass is $2.1\, M_\odot$ (blue) and $2.5\, M_\odot$ (red). In each band, the surface tension is $\Sigma = 0.5,1,2\, {\rm MeV}/{\rm fm}^2$ from top to bottom. The grey bands indicate the estimates from the merger 
GW170817 (left panel) and the (combined) interpretations from the NICER data (right panel). }
\label{fig:LamR14}
\end{center}
\end{figure*}

It is striking that the curves in Fig.\ \ref{fig:lamMKK} appear to be straight lines to a good approximation on the doubly logarithmic plot (one can check that the allowed region is roughly approximated by $M_{\rm KK} \sqrt{\lambda} \sim (3.0-3.2)\, {\rm GeV}$). In particular, lines of constant maximal mass appear to be also lines of constant tidal deformability $\Lambda_{1.4}$. To further quantify this observation, in Fig.\ \ref{fig:LamR14}  we show $\Lambda_{1.4}$ and the corresponding radius $R_{1.4}$ as a function of the 't Hooft coupling along the lines of constant maximal mass $2.1\, M_\odot$ and $2.5\, M_\odot$, i.e., moving along the red solid and red dashed lines in Fig.\ \ref{fig:lamMKK}. We see clearly that $\Lambda_{1.4}$ and $R_{1.4}$ are not exactly constant, but their variation is intriguingly small, at least in the shown range, omitting values of the 't Hooft coupling which are orders of magnitude larger or smaller than $\lambda\sim 10$, which we know is the regime where the model reproduces properties of QCD. We can thus use these plots to make approximate parameter-independent predictions of the model. For instance, the model predicts that an equation of state whose most massive star has a mass of 2.1 solar masses will produce a 1.4 solar mass star with tidal deformability $\Lambda_{1.4} \sim 230 - 350$ and radius $R_{1.4} \sim (11.4 - 13.5)\, {\rm km}$ (upper and lower limits of the blue bands shown in Fig.\ \ref{fig:LamR14}). Interestingly, this radius prediction is in very close agreement with the range given in one of the two interpretations of the NICER data \cite{Riley:2019yda}. 
Since we know from astrophysical observations that $2.1\, M_\odot$ is the lower limit for the maximal mass, the plot also demonstrates that the lower boundaries of the blue bands are absolute lower bounds predicted by our model. It is interesting to note that the lower bound for the tidal deformability, $\Lambda_{1.4} \gtrsim 230$, is identical to the lower bound found in the V-QCD approach \cite{Jokela:2020piw}. The figure also illustrates once more the tension of producing stars above 2.5 solar masses: as the left panel shows, it is possible to produce them in a certain regime in $\lambda$, but with values for $\Lambda_{1.4}$ at the very upper end of the allowed window  and only with sufficiently large, perhaps unrealistic, values of the surface tension.

\section{Summary and outlook}
\label{sec:summary}

We have constructed neutron stars from the holographic Witten-Sakai-Sugimoto model. Our approach employs the same holographic formalism from the densest matter in the center of the star up to the low-density surface. This includes a construction of the crust, which had not been done before within holography. Our holographic nuclear matter is based on a spatially homogeneous ansatz for the non-abelian gauge fields in the bulk, which yields an approximation for a many-baryon system that is expected to work best at  high densities. In particular, we have used a recent extension of this approach to isospin-asymmetric matter. After adding a non-interacting lepton gas to our holographic setup, this allows us to incorporate equilibrium with respect to the electroweak interaction and electric charge neutrality. The crust is then constructed in the Wigner-Seitz approximation as a mixed phase of the lepton gas and nuclear matter, assuming sharp, step-like interfaces between the two phases. The comparison of the free energies of pure and mixed phases is used to determine the crust-core transition dynamically. 

After showing that a simple pointlike approximation of the holographic baryons 
is not in agreement with astrophysical data, we have computed mass-radius relations and tidal deformabilities of the neutron stars constructed from the homogeneous ansatz. Our calculation is performed in the confined geometry of the model with antipodal separation of the flavor branes. In this scenario, the model only has two parameters, the 't Hooft coupling $\lambda$ and the Kaluza-Klein mass $M_{\rm KK}$, supplemented in our approach by the surface tension of nuclear matter, which we treat as an external parameter. 
We have shown that using a realistic surface tension the model parameters can be chosen to meet the known astrophysical constraints on maximal mass, tidal deformability and radius of the star. 

We have systematically studied the parameter space to compute the astrophysically allowed window in the $\lambda$-$M_{\rm KK}$ plane.
In this simple version of the model, the corresponding parameters reproduce approximately, but not exactly, the vacuum properties used in the original works as a fit. This tension in fitting to different properties is already apparent if the saturation properties of nuclear matter are taken into account. We have also extracted 
parameter-independent predictions of the model. For instance, we have shown that any parameter set that produces a maximum mass of 2.1 solar masses or higher produces a radius for a 1.4 solar mass star larger than $11.4\, {\rm km}$ and a tidal deformability larger than $230$. 
More, and more stringent, bounds can be obtained by a more exhaustive analysis, also making use of the NICER data for the radius of a 2.1-solar-mass star. This was done in our follow-up study \cite{Kovensky:2021wzu}.

In constructing the entire neutron star from a single model our work goes beyond previous holographic -- and most traditional -- approaches. Nevertheless, several improvements are desirable.
Most notably, as discussed in the main part of the paper, we know that our approach shows large-$N_c$ artifacts in the isospin spectrum. While we can define neutron and proton number and assign electric charges to them, additional states in the continuous isospin spectrum unavoidably become relevant in the simple approximation used here. A very large symmetry energy and a resulting large proton fraction are manifestations of this artifact and an important step would be to improve the treatment of isospin, for instance by including the quantization in the bulk of the holographic setup \cite{Hata:2007mb}, which however is difficult to generalize to dense matter. Moreover, our construction of the crust is simplistic in the sense that we did not include an inner crust, which contains a mixture of near-symmetric nuclear matter and a pure neutron fluid. One can try to construct such an inner crust within the setup used here. Other extensions would be to compute the surface tension dynamically within the given model, or to include 
pion condensation, which may coexist with nuclear matter in the relevant part of the phase diagram within the present approximation \cite{Kovensky:2021ddl}. It is also possible to extend our approach to include strangeness and check whether hyperons appear in the stars constructed from our holographic model. Also, one can straightforwardly -- but with more numerical effort -- extend our study to the deconfined geometry of the model. The advantage would be that a nontrivial temperature dependence can be included, relevant for the simulation of neutron star mergers, and that the transition to a chirally restored phase could be included. This would be interesting for a hybrid star containing quark matter (or quarkyonic matter \cite{Kovensky:2020xif}) in the core. However, in currently employed approximations the chiral phase transition occurs at relatively large densities, perhaps too large to be relevant for the interior of neutron stars. One could also consider computing transport properties of our neutron star matter. Finally, it would be interesting to see if our construction of the holographic crust is also applicable to similar holographic models.

\vspace{-0.5cm}

\begin{acknowledgments}
We thank Nils Andersson, Christian Ecker, Fabian Gittins, Matti J\"{a}rvinen, and Anton Rebhan for valuable comments and discussions. In the early stages, this work was supported by the Leverhulme Trust under grant no RPG-2018-153. A.P.\ is supported by an Engineering and Physical Sciences Research Council \mbox{(EPSRC)}
Mathematical Sciences Fellowship at the University of
Southampton. The work of N.K.\ is supported by the ERC Consolidator Grant 772408-Stringlandscape.
\end{acknowledgments}

\bibliography{references}

\end{document}